\documentclass[12pt]{iopart}
\usepackage{iopams}
\usepackage{setstack}
\usepackage{titlesec}
\titleformat{\section}{\bfseries \sffamily}{\thesection. }{0.1em}{}
\titleformat{\subsection}{\bfseries \sffamily \footnotesize}{\thesubsection.}{0.5em}{}
\usepackage{tocloft}
 \usepackage{hyperref}
\hypersetup{colorlinks, linkcolor=blue, citecolor=blue, urlcolor=black}

\usepackage[linesnumbered,ruled,vlined,nofillcomment]{algorithm2e}
\usepackage{graphicx}
\usepackage[caption=false]{subfig}
\usepackage{amssymb, amsthm, amsopn}

\DeclareMathOperator*{\argmax}{arg\,max}
\DeclareMathOperator*{\argmin}{arg\,min}

\makeatletter
\def\@fnsymbol#1{\ifcase#1\or ^{\dagger}\or ^{\S}\or ^{\S\S}\or ^{\P}\or ^{**}\or ^{\tsty *} \else\@ctrerr\fi\relax}
\makeatother

\DeclareGraphicsExtensions{.eps, .pdf}

\begin{document}

\title[\sffamily A divisive spectral method for network community detection]{\Huge \bfseries \sffamily A divisive spectral method for network community detection}

\author{\large \sffamily Jianjun Cheng\textsuperscript{1}, Longjie Li\textsuperscript{1}, Mingwei Leng\textsuperscript{2}, Weiguo Lu\textsuperscript{3}, Yukai Yao\textsuperscript{1}, and Xiaoyun Chen\textsuperscript{1}\footnote[6]{Corresponding author.}}
  \address{\textsuperscript{1}School of Information Science and Engineering, Lanzhou University, Gansu Province, China}
  \address{\textsuperscript{2}School of Mathematics and Computer Science, Shangrao Normal University, Jiangxi Province, China}
  \address{\textsuperscript{3} International office, Lanzhou University, Gansu Province, China}

  \ead{chengjianjun@lzu.edu.cn; chenxy@lzu.edu.cn.}

\renewcommand\abstractname{\sffamily Abstract}
\begin{abstract}
 Community detection is a fundamental problem in the domain of complex-network analysis. It has received great attention, and many community detection methods have been proposed in the last decade. In this paper, we propose a divisive spectral method for identifying community structures from networks, which utilizes a sparsification operation to pre-process the networks first, and then uses a repeated bisection spectral algorithm to partition the networks into communities. The sparsification operation makes the community boundaries more clearer and more sharper, so that the repeated spectral bisection algorithm extract high-quality community structures accurately from the sparsified networks. Experiments show that the combination of network sparsification and spectral bisection algorithm is highly successful, the proposed method is more effective in detecting community structures from networks than the others.
\end{abstract}

\section{Introduction}
Many systems can be modeled as complex networks, in which vertices represent individuals and edges describe connections between them. A significant characteristic occurred in many networks is the so-called ``community structure'', the tendency of vertices that can be partitioned into groups naturally, with denser connections between vertices within groups and sparser edges across groups \cite{Girvan_Newman:PNAS:2002, Newman_Girvan:PhRvE:2004_GN}. The communities can be groups of Web pages sharing the same topics in WWW networks \cite{Kleinberg:Science:2001, Pan_Liu:PhyA:2010}, or pathways in metabolic networks, or complexes in protein-protein interaction networks \cite{GuimeraNunes:Nature:2005, Ravasz:Science:2002, Lewis:BMC_Systems_Biology:2010, Satuluri:2011:LGS:1989323.1989399, LangeReusVan:FrontiersinComputationalNeuroscience:2014}. 

Identifying the community structures from networks is very important, because such structures can have significant influences on the function of networks. Therefore, there have been considerable researches on the problem of community detection, and a large number of methods have been developed and applied to various networks. In this paper, we focus on spectral methods, especially on bisection spectral methods, for their sound theoretical principles. The spectral methods are originated as a kind of graph-partitioning methods \cite{DonathHoffman:IBMJ.Res.Dev.:1973, Fiedler:CzechoslovakMathematicalJournal:1973, PothenSimonLiou:SIAMJournalonMatrixAnalysisandApplications:1990, SpielmanTeng:InIEEESymposiumonFoundationsofComputerScience:1996, ChanP.Szeto:SIAMJournalonScientificComputing:1997}, then they developed into a kind of classical methods for clustering \cite{ShiMalik:IEEETrans.PatternAnal.Mach.Intell.:2000, NG:NIPS2001_2092:2002, MeilaShi:AISTATS:2001, Luxburg:StatisticsandComputing:2007} and for community detection \cite{DonettiMunoz:JournalofStatisticalMechanics:TheoryandExperiment:2004, ArenasDiaz-GuileraPerez-Vicente:Phys.Rev.Lett.:2006, ChengShen:JournalofStatisticalMechanics:TheoryandExperiment:2010, Newman:Phys.Rev.E:2013, LangeReusVan:FrontiersinComputationalNeuroscience:2014, ChauhanGirvanOtt:Phys.Rev.E:2009, Newman:PhRvE:2006, ShenChengFang:Phys.Rev.E:2010, ShenCheng:JournalofStatisticalMechanics:TheoryandExperiment:2010, Nascimento:DiscreteAppl.Math.:2014, CapocciServedioCaldarelliColaiori:PhysicaA:StatisticalandTheoreticalPhysics:2005, GennipHunterAhnElliottLuhHalvorsonReidValasikWoTitaBertozziBrantingham:SIAMJournalofAppliedMathematics:2013} in the fields of data mining and complex network analysis, separately. 

For community detection, the spectral methods utilizes the eigenspectra of various types of network-associated matrix to identify the community structure. For instance, through analyzing the spectrum of the network Laplacian matrix, Donetti \textit{et~al.} \cite{DonettiMunoz:JournalofStatisticalMechanics:TheoryandExperiment:2004} projected the network vertices into a tunable-dimensionality eigenvector space, and the community structure corresponding to the global maximum of \textit{modularity} \cite{Newman_Girvan:PhRvE:2004_GN} over all possible dimensions of the eigenvector spaces was found finally. Arenas \textit{et~al.} \cite{ArenasDiaz-GuileraPerez-Vicente:Phys.Rev.Lett.:2006} reported the existence of a connection between the spectral information of the Laplacian matrix and the hierarchical process of emergence of communities at different time scales, which can be utilized to extract community structures from networks. Based on the normalized Laplacian matrix and its eigenvalues, Chen \textit{et~al.} \cite{ChengShen:JournalofStatisticalMechanics:TheoryandExperiment:2010} demonstrated that the stable local equilibrium states of the diffusion process can reveal the inherent community structures of networks, which can be extracted through optimizing the conductance of networks directly. Newman \cite{Newman:Phys.Rev.E:2013} discussed the equivalence between community detection and the normalized-cut graph partitioning, and gave spectral algorithms based on the normalized Laplacian matrix of networks to solve the two types of problems. Lange \textit{et~al.} \cite{LangeReusVan:FrontiersinComputationalNeuroscience:2014} examined the spectra of normalized Laplacian matrix of the macroscopic anatomical neural networks of the macaque and cat, and of the microscopic network of the C.elegans, and revealed an integrative community structure in these neural networks.

In addition to the Laplacian matrix and the normalized Laplacian matrix, the eigenspectra of other types of network-associated matrix were used to extract community structures as well. For example, Chauhan \textit{et~al.} \cite{ChauhanGirvanOtt:Phys.Rev.E:2009} found that the spectrum of the network adjacency matrix has some eigenvalues that are significantly larger than the magnitude of the rest of the eigenvalues, which indicated the number of communities in the network. Newman \cite{Newman:PhRvE:2006} divided the network vertices into two groups according to the signs of elements of the leading eigenvector of the ``modularity matrix'' first and then subdivided those groups based on the ``generalized modularity matrix'' recursively. Shen \textit{et~al.} \cite{ShenChengFang:Phys.Rev.E:2010} based on the network covariance matrix to uncover the multiscale community structure, and defined a ``correlation matrix'' to extract the multiscale community strucutre from the heterogeneous network utlizing its eigenvectors. And in Ref. \cite{ShenCheng:JournalofStatisticalMechanics:TheoryandExperiment:2010}, Shen \textit{et~al.} found that the normalized Laplacian matrix and the correlation matrix outperform the other three types of aforementioned matrix in detecting community structures from networks. To overcome the resolution limit problem of modularity, Nascimento \cite{Nascimento:DiscreteAppl.Math.:2014} constructed a new network based on the leading eigenvectors of those ``clustering coefficient matrix'' calculated for every vertex to extract the final community structure. Capocci \textit{et~al.} \cite{CapocciServedioCaldarelliColaiori:PhysicaA:StatisticalandTheoreticalPhysics:2005} utilized the first few eigenvectors of the network transition matrix to calculate the correlations between vertices to determine whether they belong to the same community or not. Gennip \textit{et~al.} \cite{GennipHunterAhnElliottLuhHalvorsonReidValasikWoTitaBertozziBrantingham:SIAMJournalofAppliedMathematics:2013} exploited a standard spectral clustering algorithm based on the transition matrix to identify social communities among gang members in the Hollenbeck policing district in Los Angeles. %

Among all these spectral methods, the bisection spectral methods are a special scenario. They divided the network into two parts utilizing some information of a certain eigenvector, such as the \textit{median} value of the eigenvector components corresponding to the smallest non-zero eigenvalue of the Laplacian matrix for graph partitioning \cite{PothenSimonLiou:SIAMJournalonMatrixAnalysisandApplications:1990, ChanP.Szeto:SIAMJournalonScientificComputing:1997}, the signs of components of the leading eigenvector of the (generalized) modularity matrix \cite{Newman:PhRvE:2006}, or the signs of the elements of the eigenvector corresponding to the second largest eigenvalue of the normalized Laplacian matrix \cite{Newman:Phys.Rev.E:2013} for community detection. All these literatures have derived the mathematical formulas as a support, hence benefited from the solid mathematical foundations, the results acquired by the bisection spectral methods are more interpretable, more credible and more persuasive than those based only on experiences or on empirical studies.

When used in applications of traditional graph partitioning, such as VLSI circuit design, load balance or communication reduction in parallel computing, \textit{etc.}, the bisection spectral methods tended to partition the network into equal-sized subgraphs. For community detection, we need to obtain a community structure as natural as possible. For a two-community network, the bisection spectral methods can partition it into two parts corresponding to the community structure successfully \cite{Newman:Phys.Rev.E:2013}. However, in general cases, the network contains more than two communities. For those networks, a natural idea is to bisect the two subgraphs recursively after the first division, but it is not guaranteed to acquire the most natural community structure. That is the reason why Newman \cite{Newman:PhRvE:2006} subdivided the subgraphs based on the generalized modularity matrix after the first division rather than bisecting recursively based on the leading eigenvector of modularity matrix only. Even so, the result is not ideal. So Newman had to employ a vertex-moving strategy to fine-tune the communities after each division. The communities extracted by this method are, by definition, indivisible subgraphs, which are always too trivial in many networks to be acceptable, and the extracted community structure often deviates far from the ground truth. For this reason, Newman \cite{Newman:Phys.Rev.E:2013} pointed out that how to generalize the bisection spectral methods to networks containing more than two communities is still an open problem.

In this paper, we propose a method to solve the problem. We observed that from several networks with apparent community structure, in which communities are separated clearly and sharply, the recursive bisection spectral method can extract the high-quality community structure definitely. Inspired by the observation, we propose a network-sparsification algorithm to promote the prominence of the community structure through removing some edges from the network. And then we propose a repeated bisection algorithm to extract the community structure from the sparsified network.

The remainder of this paper is organized as follows. In section \ref{observation}, we demonstrate the observation mentioned above using an example network with apparent community structure, the proposed method is elucidated in section \ref{proposal}, the experimental results is shown in section \ref{experiments}, and this paper is ended with a conclusion in section \ref{conclusion} .

\section{Observation\label{observation}}
\begin{figure*}[htbp]
\centering
\subfloat[\label{fig:observation:ground_truth}]{\includegraphics[scale=0.42]{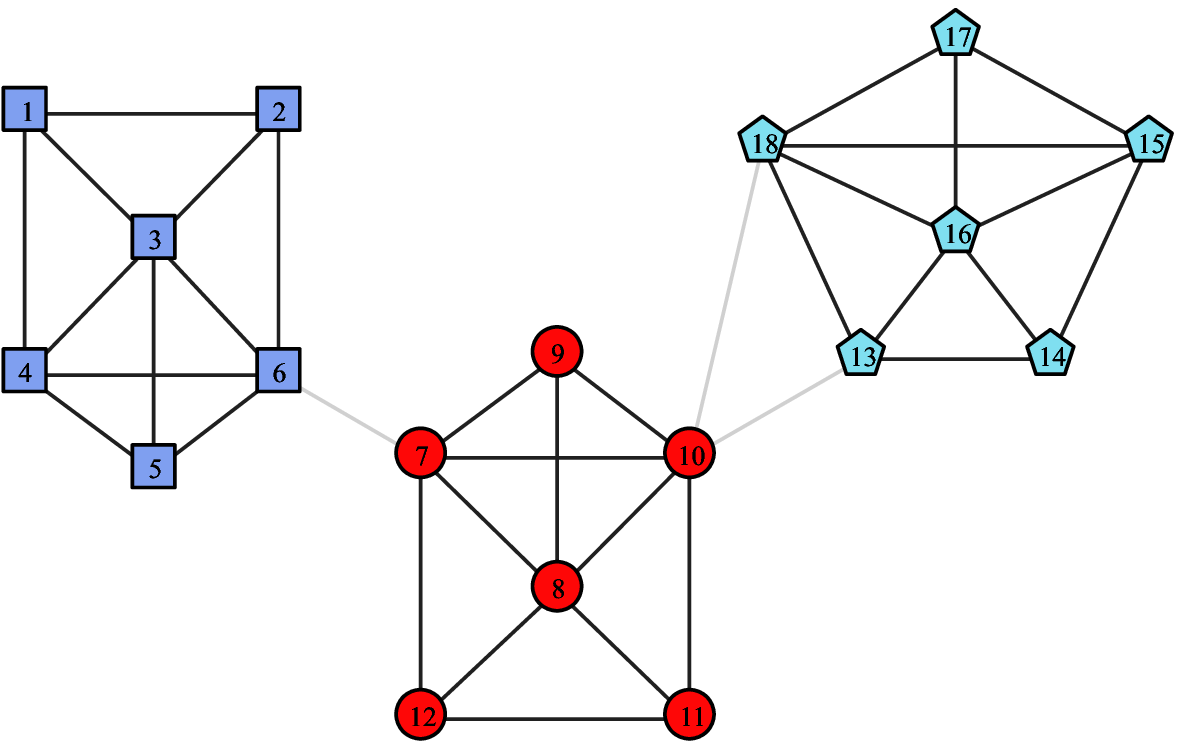}}\ {}
\subfloat[\label{fig:observation:step1}]{\includegraphics[scale=0.42]{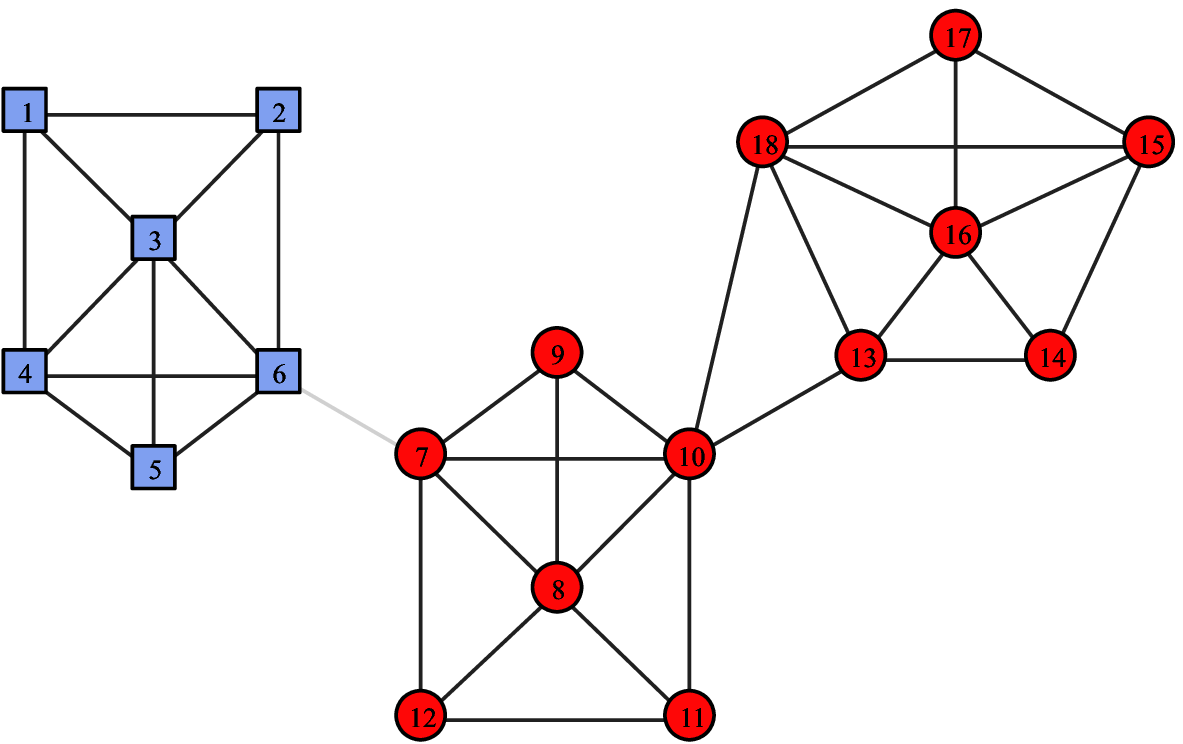}}\ {}
\subfloat[\label{fig:observation:step2}]{\includegraphics[scale=0.42]{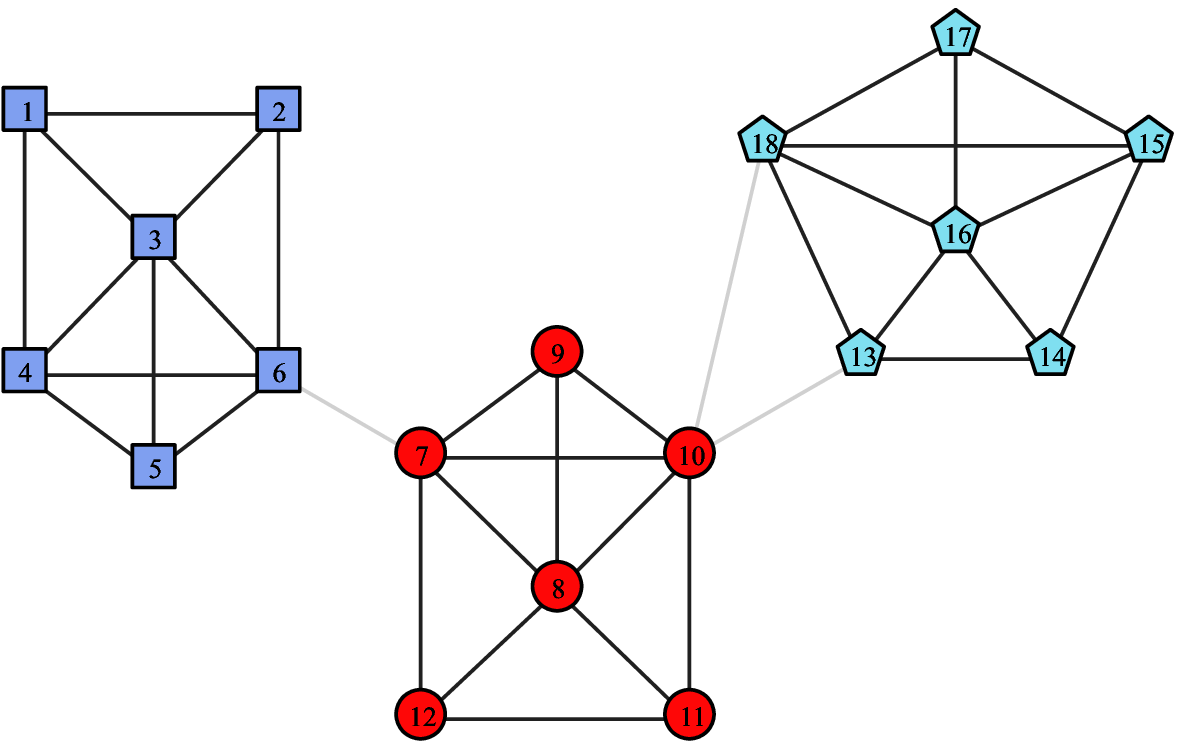}}
\caption{\small \textbf{A simple network containing more than two communities.} \protect\subref{fig:observation:ground_truth} The ground-truth community structure; \protect\subref{fig:observation:step1} The community structure corresponding to the first bisection; \protect \subref{fig:observation:step2} The community structure corresponding to the second bisection. The different vertex shapes and shades indicate different communities, the black lines represent edges within communities, and the light gray lines represent connections across communities. This illustration style also applies to the next figures.}
\end{figure*}

Although the recursive bisection spectral method is not guaranteed to obtain the best community structures in general cases, we have observed that it does work well on some special networks. For example, the simple network illustrated in Figure \ref{fig:observation:ground_truth} is a such special network, which contains 3 communities, and the community boundaries are evident. Applying the recursive bisection spectral method to this network can get the ideal result, Figure \ref{fig:observation:step1} shows the result of the first bisection. Bisecting recursively the two subgraphs in Figure \ref{fig:observation:step1}, we obtain the resulting community structure presented in Figure \ref{fig:observation:step2}. Obviously, it is identical to the ground-truth community structure.

\captionsetup[subfigure]{subrefformat=subparens, labelformat=simple, listofformat=subsimple}
\renewcommand{\thesubfigure}{(\alph{subfigure})}
\renewcommand{\figurename}{\bfseries \sffamily \small Figure}

In fact, we have also tested the recursive bisection spectral method on some other networks that have the similar characteristics as the one illustrated in Figure \ref{fig:observation:ground_truth}, we observed that all results are satisfactory. That is to say, applying recursive bisection spectral method to networks, in which communities are well defined and separated clearly and sharply, can extract high-quality community structures.

Inspired by this observation, we propose a method in this paper to extend the recursive bisection spectral method to a repeated bisection spectral method that can deal with networks which contain more than two communities and the community boundaries are not so sharp. We first remove some edges from the network to make the community boundaries more clearer and more sharper, then use the repeated bisection spectral method to extract the final community structure.

\section{The proposed method\label{proposal}}
The proposed method is comprised of two algorithms. The first one is responsible for sparsifying the network to promote the prominence of the community structure by removing some edges from the network, the second one is the repeated bisection spectral algorithm to extract the community structure accurately from the sparsified network.

Facilitating the description of the proposed method, some notations are given in definition form as follows.

\theoremstyle{definition}
\newtheorem{Def}{Definition}
\begin{Def}
 A network is an unweighted and undirected simple graph $G=(V,E)$, where $V$ and $E$ are the vertex set and the edge set, respectively, and $|V|=n$, $|E|=m$.
\end{Def}

\begin{Def}
A community structure of network $G$ is a partition of the network, denoted as $\mathcal{\boldsymbol{CS}}=\{C_1,C_2,\cdots, C_\mathcal{K}\}$, where $C_i\subseteq V$, $\cup_{i=1}^\mathcal{K}C_i=V$ and $C_i\cap C_j=\phi$ ($i, j=1,2,\cdots, \boldsymbol{\mathcal{K}}$, and $i\ne j$), and $\boldsymbol{\mathcal{K}}$ is the number of communities in the partition. In accordance with the concept of community, an additional constraint, $\sum_{i=1}^\mathcal{K}\Big|\{(u,v)|(u,v)\in E, u\in C_i, v \in C_i\}\Big| >> \sum_{i,j=1}^{\mathcal{K}}\Big| \{(u,v)|(u,v)\in E, u \in C_i, v\in C_j, i\ne j\}\Big|$ is always attached to the partition, which means that the edges within communities are much denser than those across different communities.
\end{Def}

\begin{Def}
$N(v)$ is the neighbour set of vertex $v$, i.e., $N(v)=\{u|(v,u)\in E\}$
\end{Def}

\begin{Def}
$d_v$ is the degree of vertex $v$, it is the number of edges incident to vertex $v$, i.e., $d_v=| N(v)|$
\end{Def}

\begin{Def}
The similarity between a pair of vertices, $u$ and $v$, is denoted as $Sim(u,v)$.
\end{Def}

\subsection{Network sparsification}
The object of the network-sparsification algorithm is to make the community boundaries more clearer and more sharper by removing some edges from the network, but which edges should be removed to reach the goal? The best answer is the edges across communities certainly, but that is obviously the ideal scenario because we cannot determine which edge across communities conveniently, or the community structure can be extracted easily. 

However, according to the concept of community, edges within communities are much denser than those across communities, that means every vertex and most of its neighbours should belong to the same community. Therefore, if we use a neighbour- related measure to calculate the similarity, $Sim(u,v)$, between any pair of vertices, $u$ and $v$, connected by an edge, the similarities between vertices in the same community will generally and intuitively be much larger than the counterparts between vertices located in different communities. 

Based on this idea, we employ a very simple strategy to sparsify the network. First, we define the similarity between any pair of vertices, $u$ and $v$, as follows,
\[Sim(u,v)=\left\{\begin{array}{cl}
  \displaystyle\frac{|N(u)\cap N(v)|}{d_u} & (u,v) \in E\\
  \displaystyle 0 & \text{otherwise}
  \end{array}\right . .\]
Obviously, $Sim(u,v)\neq Sim(v,u)$ in general cases, i.e., this similarity is asymmetric.

Then, we remove the edges which connect pairs of vertices that the similarity between them are smaller than a given threshold, $\theta$, from the network, but for edges that the degree of any one of end vertices is not larger than 3, we give special consideration. The entire procedure is listed as Algorithm \ref{alg:network_sparsification}. 

\addtolength{\intextsep}{-1ex}
\SetAlCapFnt{\footnotesize}
\SetAlFnt{\footnotesize}
\begin{algorithm}[htbp]
 \DontPrintSemicolon
 \caption{\label{alg:network_sparsification}The network-saparsification algorithm}
 \KwIn{$\boldsymbol{G(V,E)}$, network; $\boldsymbol{\theta}$, similarity threshold}
 \KwOut{$\boldsymbol{G}$, sparsified network}

 \BlankLine
 \ForEach{$(u,v)\in E$}{
    $dmin\leftarrow \min\{d_u,d_v\}$\;
    $x \leftarrow \argmin_w\left\{d_w|w\in\{u,v\}\right\}$\;
    \If{$dmin\leqslant 2$}{continue\;}
    \If{$dmin=3$}{
     \If{$max\{d_w|w\in N(x)\}\leqslant dmin$}{continue\;}
    }
   \BlankLine
   \If{$(Sim(u,v)<\boldsymbol{\theta})$ and $(Sim(v,u)<\boldsymbol{\theta})$}{
     $\boldsymbol{G}.$remove\_edge$(u,v)$\;
   }
 }
 \BlankLine
 \Return $\boldsymbol{G}$
\end{algorithm}

The operations are almost self-explanatory. For each edge in the network, if the degree of any one of end vertices is not larger than 2, we bypass this edge directly. For the edge that the degree of one of end vertices is equal to 3, we determine whether there exists any vertex whose degree is larger than that end vertex in its neighbours or not. If no, this edge is also neglected. The aim of these special consideration is to keep the network from being partitioned into trivial or even single-vertex communities in the network-sparsification stage. For each of other edges in the network, we calculate two similarities between two end vertices, if both values of the two asymmetric similarities are smaller than the given threshold, $\theta$, we remove this edge from the network. And finally, the sparsified network is returned.

\subsection{Repeated bisection spectral algorithm}
After sparsification, we extract the community structure from the sparsified network using our proposed bisection spectral algorithm. Our proposal is a repeated bisection spectral algorithm, it is based on the signs of elements of the eigenvector corresponding to the second largest eigenvalue of the network transition matrix.

In Ref. \cite{Newman:Phys.Rev.E:2013}, starting from optimizing modularity, Newman derived the formulas to describe the rationale of his bisection spectral method (although, it can be fit for two-community networks only), and achieved a formula for the modularity
\[Q=\frac{\lambda}{2}, \]
where $\lambda$ is the eigenvalue of the generalized eigenvector equation
\begin{equation}\label{eq:generalized_eigenvector_equation}
\boldsymbol{As}=\lambda \boldsymbol{Ds}
\end{equation}
with constraint $\boldsymbol{K^Ts}=0$. Where $\boldsymbol{A}$ is the network adjacency matrix, $\boldsymbol{D}$ is a diagonal matrix with elements equal to the vertex degrees, i.e., $D_{ii}=d_i$, $\boldsymbol{K}$ is a vector with elements $k_i=d_i$, and eigenvector $\boldsymbol{s}$ is the solution vector whose elements equal to $\pm 1$, i.e., $s_i=+1$ indicates to put vertex $i$ into group 1, and $s_i=-1$, into group 2.

Our proposed repeated bisection spectral algorithm is also based on the above formulas. Let us consider first the scenario bisecting a network in two parts, and then call it repeatedly. To maximize the value of the modularity $Q$, we should choose $\lambda$ to be the largest eigenvalue of Eq. (\ref{eq:generalized_eigenvector_equation}), but it is impossible here. Because it is obvious that vector $\boldsymbol{s}=\boldsymbol{1}=(1,1,\cdots,1)^T$ is an eigenvector of Eq. (\ref{eq:generalized_eigenvector_equation}), and according to the Perron-Frobenius theorem, it must corresponds to the largest (most positive) eigenvalue, but $\boldsymbol{s}=\boldsymbol{1}=(1,1,\cdots,1)^T$ fails to satisfy constraint $\boldsymbol{K^Ts}=0$. Therefore, what we can do best is to choose $\lambda$ to be the second largest eigenvalue to maximize the modularity, $Q$, and to choose the corresponding eigenvector to be the solution vector $\boldsymbol{s}$. However, this eigenvector is a real-number vector, considering the constraint of $\boldsymbol{s}$, $s_i=\pm 1$, we can simply round the value of $s_i$ to $\pm 1$ to get the solution vector. This operation is equivalent to checking the signs of elements of $\boldsymbol{s}$ to put the corresponding vertices into group 1, or into group 2. Hereafter in this paper, we call ``the eigenvector corresponding to the second largest eigenvalue'' as ``the second eigenvector'' to facilitate the description.

\setcounter{footnote}{0}
To solve Eq. (\ref{eq:generalized_eigenvector_equation}), we simply rearrange its terms, and obtain that\footnote{The matrix $\boldsymbol{D}$ is invertible because all subnetworks involved are connected.}
\begin{equation}
\boldsymbol{D^{-1}As}=\lambda \boldsymbol{s}.
\end{equation}
The matrix
\begin{equation}
\boldsymbol{T}=\boldsymbol{D^{-1}A}
\end{equation}
is the transition matrix corresponding to random walk in the network,
our proposed algorithm is based on it:
for the sparsified network, we compute the second eigenvector of the transition matrix, and then divide the vertices of the sparsified network into two communities according to the signs of the second eigenvector elements. This is a bisection operation that divide the network vertices into two communities only, to extract the resulting community structure containing multiple communities, we construct a subnetwork for each community, and from all subnetworks, the one whose split can lead to a new community structure with the maximal modularity is selected to perform the bisection division really. This division operation is repeated until the community number reaches the given number of communities, $\boldsymbol{\mathcal{K}}$.

The pseudo code outlining the entire procedure is listed in Algorithm \ref{alg:community_detection:main}. After sparsification, the network itself might become disconnected. We take each connected component as a community, and all of the connected components comprise the initial community structure $\boldsymbol{CS}$. Next, for each community $C_i\in\boldsymbol{CS}$, a subnetwork of $\boldsymbol{G}$, $\boldsymbol{sg_i}$, is constructed and bisected into two subgraphs afterwards by calling function ``spectra\_bisection()'', then we calculate the modularity of the new community structure corresponding to this bisection. From all bisections, the one with the maximal modu\-larity (the corresponding community is $C_j$) is selected to be accepted as the real division by removing $C_j$ from $\boldsymbol{CS}$ and inserting two obtained communities $C_{j1}$ and $C_{j2}$ into $\boldsymbol{CS}$. This operation is repeated until the number of communities reaches $\boldsymbol{\mathcal{K}}$, and we obtain the resulting community structure finally.

\addtolength{\textfloatsep}{-3ex}
\begin{algorithm}[htbp]
\DontPrintSemicolon
\LinesNotNumbered
\caption{\label{alg:community_detection:main}The repeated bisection spectral community detection algorithm}
\KwIn{$\boldsymbol{G(V,E)}$, network; $\boldsymbol{\mathcal{K}}$, number of communities in the resulting community structure}
\KwOut{$\boldsymbol{CS}$, community structure}
\BlankLine
\nl$\boldsymbol{CS}\leftarrow \boldsymbol{G}.\text{connected\_components()}$\;
\nl \While{$|\boldsymbol{CS}|<\boldsymbol{\mathcal{K}}$}{
\nl \ForEach{$C_i \in \boldsymbol{CS}$}{
\nl $\boldsymbol{sg_i}\leftarrow\boldsymbol{G}.\text{subgraph}(C_i)$\;
\nl $(C_{i1},C_{i2})\leftarrow \text{spectra\_bisection}(\boldsymbol{sg_i})$\;
\nl calculate modularity, denoted as $Q_i$, of the community structure supposing that $C_i$ is removed from $\boldsymbol{CS}$, $C_{i1}$ and $C_{i2}$ are inserted into $\boldsymbol{CS}$\;
}
\nl $j\leftarrow \argmax_i\{Q_i|i=1,2,\cdots,|\boldsymbol{CS}|\}$\;
\nl $\boldsymbol{CS}\leftarrow \boldsymbol{CS} \backslash \{C_j\}$\;
\nl $\boldsymbol{CS}\leftarrow \boldsymbol{CS} \cup \{C_{j1},C_{j2}\}$\;
}
\nl\Return $\boldsymbol{CS}$

\LinesNotNumbered
\BlankLine\BlankLine
\SetKwFunction{bsSplitName}{spectra\_bisection}
\SetKwProg{bsSplit}{Function}{}{}
\setcounter{AlgoLine}{0}
\bsSplit{\bsSplitName$\!\!(\boldsymbol{sg})$}{ 
\nl $\boldsymbol{A}\leftarrow \boldsymbol{sg}.\text{adjacency\_matrix}()$\;
\nl $\boldsymbol{D}\leftarrow diag(\sigma_i)$ \,\texttt{/*}\,\,where $\sigma_i=\sum_j^nA_{ij}, i=1,2,\cdots,n.$\,\texttt{*/}\;
\nl $\boldsymbol{T}\leftarrow \boldsymbol{D^{-1}A}$\;
\nl $(\lambda_2,\boldsymbol{x_2})\leftarrow \text{second\_largest\_eval\_evec}(\boldsymbol{T})$\;
\BlankLine
$\boldsymbol{s}\leftarrow\boldsymbol{x_2}$\;
\nl $C_1\leftarrow\{v|\boldsymbol{s}[v]>0\}$\;
\nl $C_2\leftarrow\{v|\boldsymbol{s}[v]<0\}$\;
\BlankLine
\nl \Return{$(C_1,C_2)$}\;
}
\end{algorithm}

In Algorithm \ref{alg:community_detection:main}, the function ``spectra\_bisection()'' is responsible for the bisection operation of the network/subnetwork, $\boldsymbol{sg}$. In this function, the second largest eigenvalue $\lambda_2$ and the corresponding eigenvector $\boldsymbol{x_2}$ of transition matrix $\boldsymbol{T}$ are computed first. Then, $\boldsymbol{x_2}$ is taken as the solution vector $\boldsymbol{s}$, and the vertices corresponding to the positive elements and the negative elements of $\boldsymbol{s}$ are put into group $C_1$ and group $C_2$, respectively. At last, the tuple of the two groups, $(C_1, C_2)$, is returned as the result.

For the community number, $\boldsymbol{\mathcal{K}}$, although some strategies, including some spectral strategies \cite{ChauhanGirvanOtt:Phys.Rev.E:2009, CapocciServedioCaldarelliColaiori:PhysicaA:StatisticalandTheoreticalPhysics:2005, ShenCheng:JournalofStatisticalMechanics:TheoryandExperiment:2010}, can be used to determine its value from the network automatically. But in practice, the numbers obtained using these strategies always differ from the exact numbers of communities contained in the ground-truth community structures more or less. In fact, to our knowledge, how to determine the exact number of communities contained in a network is still a challenging problem. Therefore, we do not invest time to acquire the community number here, but take it as a parameter of our proposed algorithm, and give its value on each network directly in our experiments instead.

\subsection{Implementation techniques}

At first glance, it seems that we need to invoke the bisection for each community in current community structure by calling the function ``spectra\_bisection()'' in each iteration of the ``while'' loop in Algorithm \ref{alg:community_detection:main} , to select the community whose bisection can lead to a new community structure with the maximal modularity. However, it is obvious that a large amount of bisections are duplicated, which leads to a lower efficiency.

To implement the algorithm efficiently, rather than bisecting each community in current community structure in each iteration, we maintain a binary tree to track the entire division procedure, which is con\-structed as follows.
\begin{itemize}
	\item It begins with vertex set $V$ in the original network as its root;
	\item If the network is disconnected after sparsification, one community in the initial community structure is taken as left child of the root, the other communities of the initial community structure are taken as right child of the root. If the right child contains more than one community, we take it as a new root, one community in it as its left child, and the remainder communities as its right child to construct a subtree recursively. Figure \ref{fig:imptech:a} shows an example binary tree of a such network whose initial community structure contains 3 components after sparsification.
	\begin{figure}
	\centering
	\subfloat[]{
		\label{fig:imptech:a}
	  \raisebox{1ex}{\includegraphics[scale=0.7]{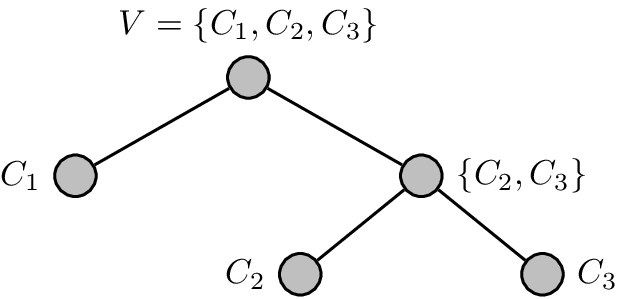}}
	}\quad
	\subfloat[]{
	  \label{fig:imptech:b}
	  \includegraphics[scale=0.7]{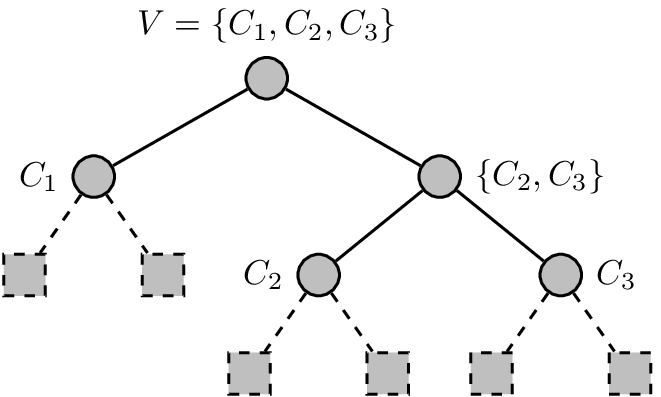}
	}\\
		\caption{\textbf{An example binary tree for the initial community structure containing 3 disconnected components, $C_1$, $C_2$, and $C_3$, after sparsification.} \protect\subref{fig:imptech:a}.~The binary tree for the initial community structure. \protect\subref{fig:imptech:b}.~The new version of the binary tree with the sentinel nodes attached. The nodes plotted in square represent the sentinel nodes of communities, each community and its sentinel nodes connect with dashed lines.}
	\subfloat[]{
	  \label{fig:imptech:c}
	  \raisebox{1.5ex}{\includegraphics[scale=0.7]{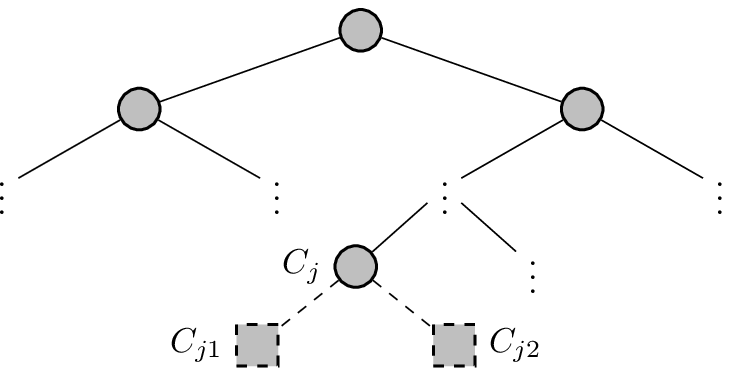}}
	}\quad
	\subfloat[]{
		\label{fig:imptech:d}
		\includegraphics[scale=0.7]{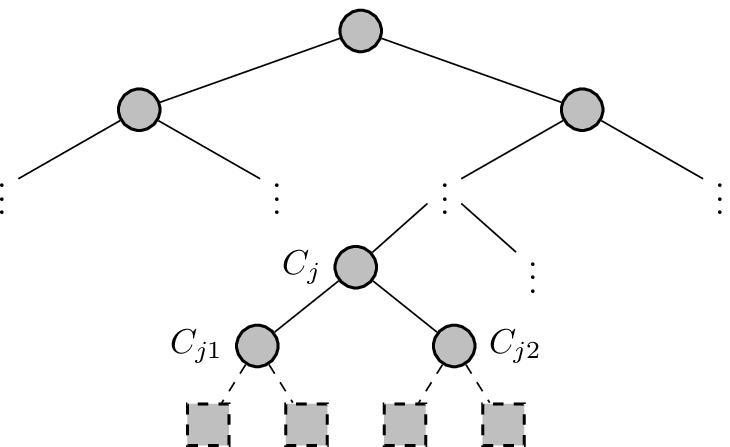}
	}
	\caption{\label{fig:imptech}\textbf{The alteration of sentinel nodes to left child and right child of the selected community.} \protect\subref{fig:imptech:c}~$C_j$ is the selected community, $C_{j1}$ and $C_{j2}$ are the two sentinel nodes of $C_j$. \protect\subref{fig:imptech:d}~$C_{j1}$ and $C_{j2}$ are altered to left child and right child of $C_j$, respectively, which means $C_j$ is removed from current community structure and $C_{j1}$ and $C_{j2}$ are inserted into it. The sentinel nodes of $C_{j1}$ and $C_{j2}$ are also plotted.}
	\end{figure}

  \item For each community in current community structure, after it is bisected for the first time when selecting the community to perform the real bisection division, we attach the two groups to the community as its sentinel child nodes. In the subsequent iterations, these two sentinel nodes are used directly instead of bisecting that community again. For instance, Figure \ref{fig:imptech:b} illustrates a new version of the binary tree shown in Figure \ref{fig:imptech:a} with sentinel nodes attached.

  \item For the selected community, to reflect the result that its bisection is accepted as the real division, its sentinels are altered to its left child and right child, respectively. Figure \ref{fig:imptech} shows an alteration example, in Figure \ref{fig:imptech:c}, $C_j$ is the selected community, $C_{j1}$ and $C_{j2}$ are two sentinels of $C_j$, which is obtained by previous bisection operation. After the bisection of $C_j$ is accepted as the real division, the status of the binary tree is as presented in Figure \ref{fig:imptech:d}. In the next iteration, the communities needed to be bisected are $C_{j1}$ and $C_{j2}$ only, not all of the communities in current community structure. The sentinels of $C_{j1}$ and $C_{j2}$ are also plotted in Figure \ref{fig:imptech:d}.
\end{itemize}

As mentioned above, the entire division procedure are tracked in this binary tree. With its aid, each community is needed to perform the bisection only once, and the current community structure consists of all of the leaf nodes (not the sentinel nodes) in each iteration. However, to locate a community in current community structure, we need to traverse a path from the root to the leaf node corresponding to that community, this traverse can be quite time consuming for large networks. To reduce the time consumption, we assign every node in the binary tree a number just as the tree is organized as a complete binary tree in logical, i.e., the number of root is $1$, and for each node $C_j$, if its number is $j$, then the numbers of its left child and right child are $2\times j$ and $2\times j+1$, respectively. Furthermore, we construct a hash table, which takes the numbers of nodes as its keys, to map the number to the position of the corresponding node in the tree. With the aid of the hash table, we can locate any community in the tree efficiently not only, but also need not to traverse the binary tree when determining whether a node, whose number is $i$, is a leaf node or not, but to check instead whether $2\times i$ or $2\times i+1$ is in the key set of the hash table or not quickly.

\section{Experiments\label{experiments}}
\subsection{Networks}
To evaluate the effectiveness of our proposed method, we conducted extensive experiments on 5 real-world networks, namely Zachary's karate club network \cite{Zachary:JournalofAnthropologicalResearch:1977, Girvan_Newman:PNAS:2002, Newman_Girvan:PhRvE:2004_GN}, Lusseau's bottlenose dolphin social network \cite{Lusseau:BehavioralEcologyandSociobiology:2003}, a map used in the popular strategy board game Risk \cite{Steinhaeuser:PatRL:2010}, a collaboration network of scientists working at the Santa Fe Institute \cite{Girvan_Newman:PNAS:2002}, and a network representing the schedule of regular season Division I American college football games for year 2000 season \cite{Girvan_Newman:PNAS:2002}. These networks are publicly available and their ground-truth community structures are already known, facilitating the verification and the validation of the proposed method, their scales are small enough alleviating the burden of interpretation and visualization of the results. Therefore, they are widely used as benchmarks for testing community detection algorithms or methods. The statistical information of them are listed in Table~\ref{tab:network:info}.
\begin{table}[htbp]
\centering
\caption{\label{tab:network:info}\textbf{The statistical information of the 5 networks used in our experiments.}}
\begin{tabular}{cccc} \hline
 network & vertices & edges & communities\\ \hline
 karate & 34 & 78 & 2\\
 dolphin & 62 & 159 & 2 \\
 Risk map & 42 & 83 & 6 \\
 scientist's collaboration & 118 & 197 & 6\\
 college football game schedule & 115 & 613 & 12\\ \hline
\end{tabular}
\end{table}

\subsection{Evaluation metrics}
To measure the strength of the extracted community structure, the modularity \cite{Newman_Girvan:PhRvE:2004_GN},  which is denoted as $Q$ and defined as:
\[Q=\sum_i^{\boldsymbol{\mathcal{K}}}(e_{ii}-a_i^2),\]
is a de facto metric at present, where $e_{ii}$ is the ratio of the edges within communities to the total edges in the network, and $a_i^2$ is the expected value of the ratio.

The modularity suffers from the so-called resolution limit problem \cite{FortunatoBarthelemy:ProceedingsoftheNationalAcademyofSciences:2007}. Therefore, we use two other metrics, namely $accuracy$ and $NMI$ (Normalized Mutual Information) \cite{Fred:ICSCCVPR:2003}, to evaluate the quality of the extracted community structure as well. The accuracy, denoted as $A$, is defined as the fraction of the vertices being classified into the correct communities to the total vertices in the network. And $NMI$ is define as:

\[NMI=\frac{%
  \displaystyle -2\sum_{i=1}^{|P|}\sum_{j=1}^{|C|}n_{ij}\log\Bigg(\frac{n_{ij}\cdot n}{n_i^P\cdot n_j^{C}}\Bigg)}%
{%
\displaystyle \sum_{i=1}^{|P|}n_i^P\log\Bigg(\frac{n_i^P}{n}\Bigg)+\sum_{j=1}^{|C|}n_j^C\log\Bigg(\frac{n_j^C}{n}\Bigg)
},
\]
where $P=\{P_1,P_2,\cdots,P_{K^\prime}\}$ and $C=\{C_1,C_2,\cdots,C_K\}$ are the extracted community structure and the ground-truth community structure, respectively, $n_i^P=|P_i|$, $n_j^C=|C_j|$, and $n_{ij}=|P_i\cap C_j|$.

Both the accuracy and $NMI$ take the ground-truth community structure as a baseline to measure how the extracted community structure approaches the ground truth, and then measure the ability of the community detection methods or algorithms. They both fall in the range $[0,1]$, larger is better.

\subsection{Comparison system and parameter settings}
Apart from being a bisection spectral method, our proposal falls in the category of divisive hierarchical methods as well. Therefore, to testify the superiority of our proposal, we ran the proposed method on the 5 networks and compared the results not only with two spectral-analysis based algorithms, namely the standard spectral clustering algorithm \cite{NG:NIPS2001_2092:2002} and the modularity-matrix based bisection spectral algorithm proposed by Newman \cite{Newman:PhRvE:2006} (abbreviated as Newman2006), but also with a novel hierarchical algorithm, Infohiermap \cite{RosvallBergstrom:PLoSONE:2011}, which identifies hierarchical community structures from networks via finding the shortest multilevel description of a random walk in networks. For the spectral clustering algorithm, its results are not deterministic, because it exploits the $K$-means algorithm to cluster the vertices, we present the result occurred most frequently in 20 runs of the algorithm here.

In addition, for the 2 two-community networks, we also made a comparison between the results of our proposed method and Newman's method described in Ref. \cite{Newman:Phys.Rev.E:2013} (shorted as Newman2013) as Newman2013 can be only applied to two-community networks. Furthermore, on all 5 networks, we compared the results of our proposal with the results extracted by the proposed repeated bisection spectral algorithm only without network sparsification to demonstrate the effectiveness of the proposed network sparsification algorithm. Hereafter, we refer the proposed method with network spar\-si\-fi\-cation as the complete version of our proposal (Algorithm \ref{alg:network_sparsification} + Algorithm \ref{alg:community_detection:main}) and the proposed repeated bisection spectral algorithm without network sparsification (Algoirthm \ref{alg:community_detection:main} only) as the lite version of our proposal, respectively.

For the proposed method, the similarity threshold $\theta$ in Algorithm \ref{alg:network_sparsification} works as a parameter to control the number of edges to be removed from the network. Its setting is crucial for the method, too large $\theta$ will filter out too many edges from the network, that may even destroy the skeleton of communities, leading to the failure of identifying them from the network; on the contrary, too small $\theta$ may lead to the result that few edges between communities are removed, so that the boundaries between communities will not be as clear as expected after sparsification. That is to say, the sparisification algorithm might not take its effect if $\theta$ is too small. After taking a sequence of values in $[0,0.6]$ as $\theta$ and 0.05 as an increment each time to carry out the experiments on each network, we concluded that $\theta=0.15$ seems to be the best setting for all 5 networks. For other networks, we suggest empirically that the mode of similarity values in $[0.1,0.2]$ be taken as the value of $\theta$.

For the parameter $\boldsymbol{\mathcal{K}}$ in Algorithm \ref{alg:community_detection:main}, which points out the number of communities in the resulting community structure, thus its value is naturally set to be the number listed in the last column in Table \ref{tab:network:info} on each network, correspondingly.

\subsection{Experimental results}
\begin{figure*}[!htbp]
\centering
\newcommand\figscale{0.44}
\subfloat[\label{fig:karate:ground_truth}]{
	\includegraphics[scale=\figscale]{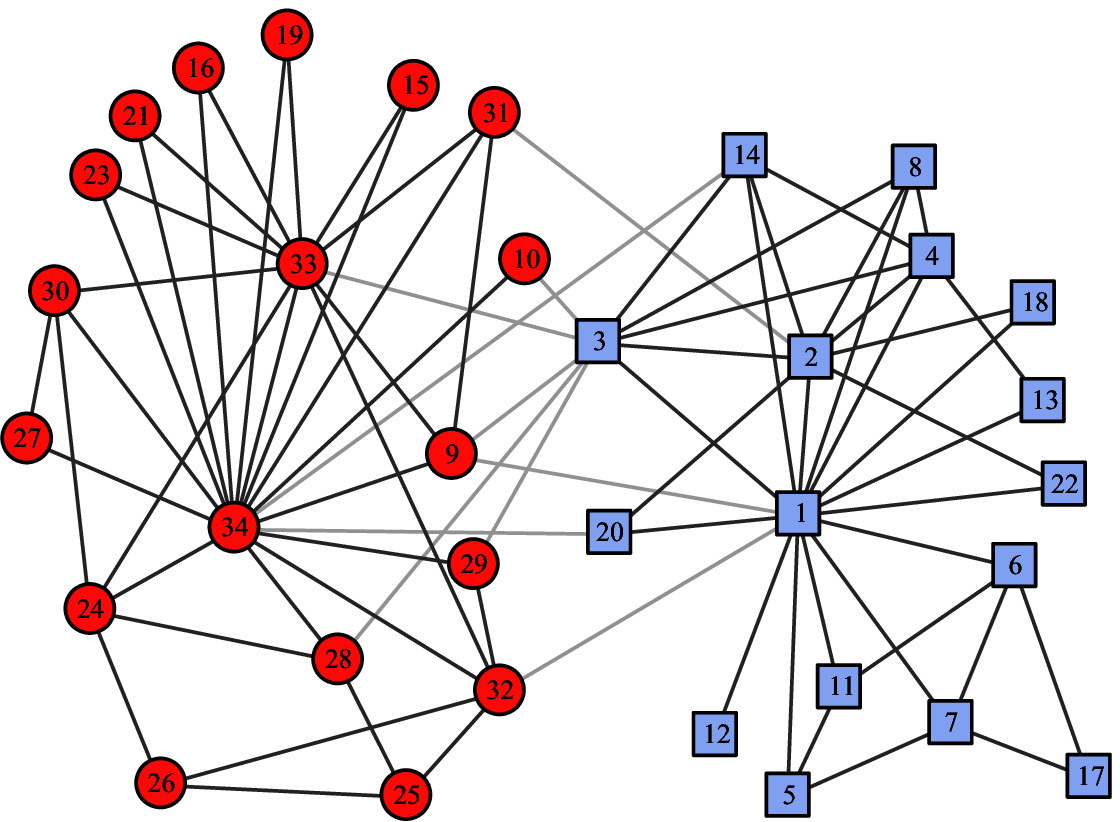}
}\\
\subfloat[\label{fig:karate:spectral_clustering}]{
	\includegraphics[scale=\figscale]{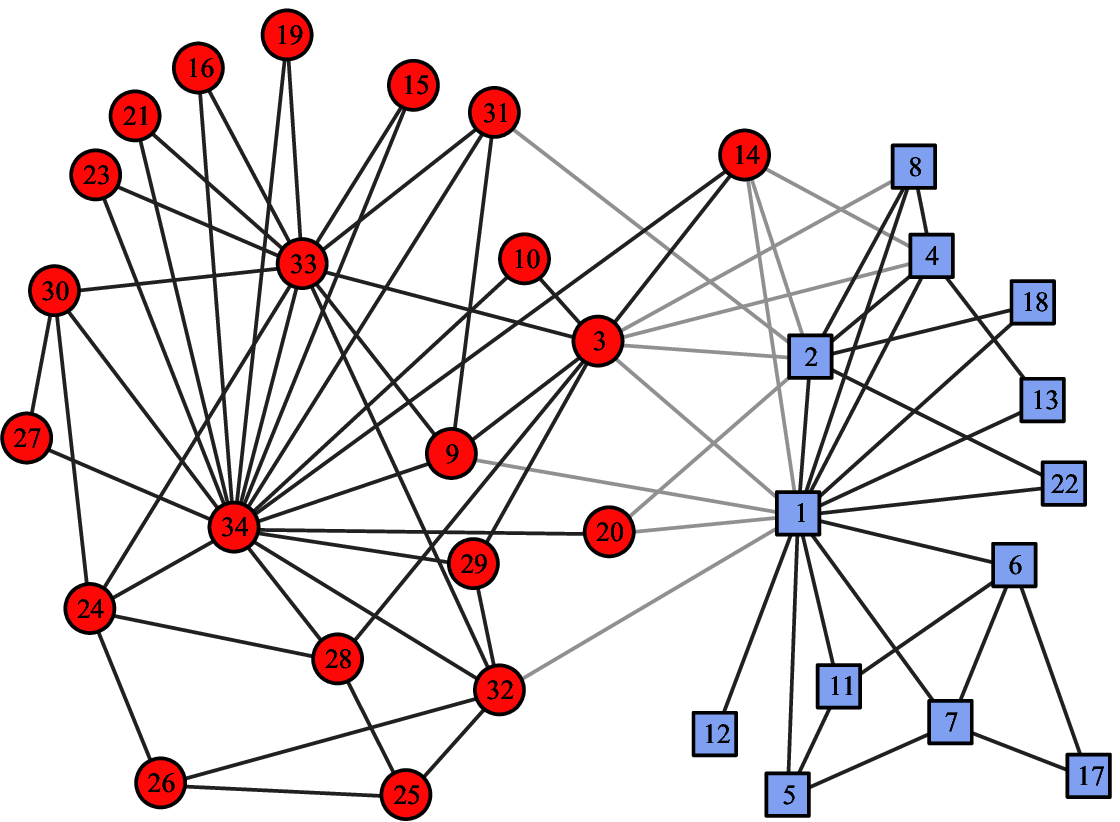}
}
\subfloat[\label{fig:karate:newman2006}]{
	\includegraphics[scale=\figscale]{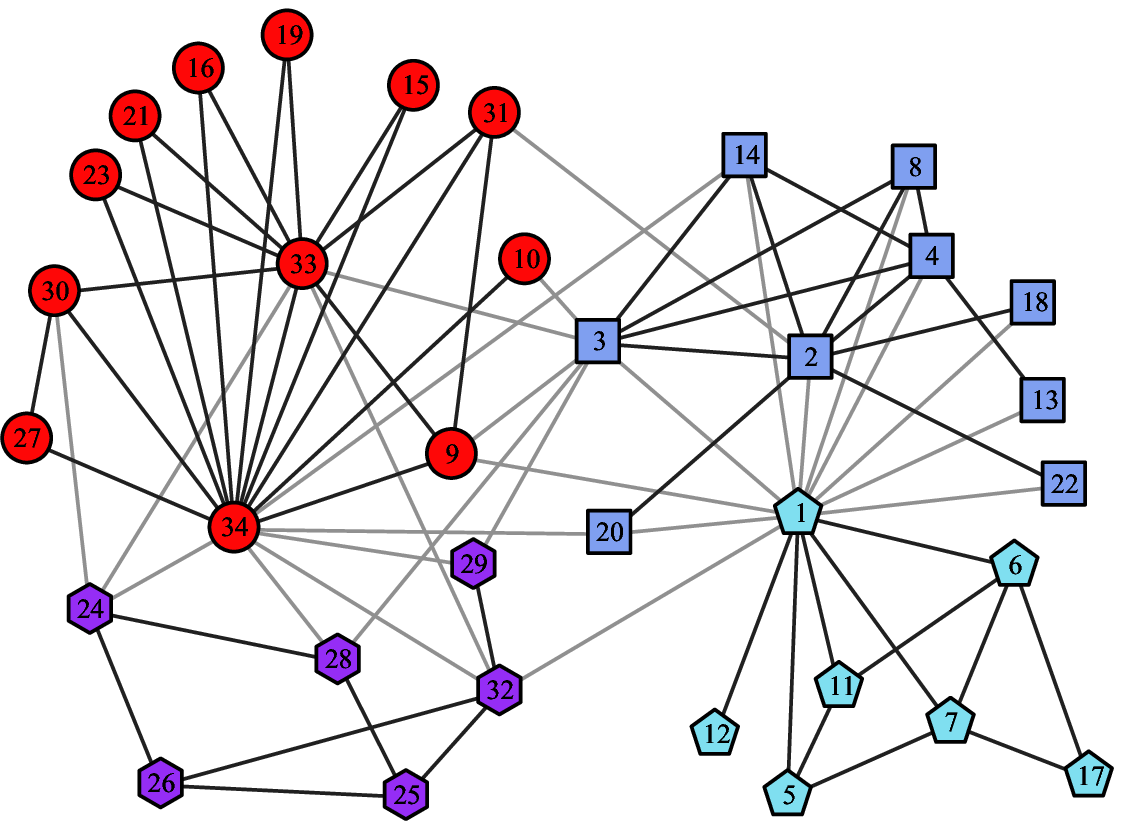}
}
\subfloat[\label{fig:karate:infohiermap}]{
	\includegraphics[scale=\figscale]{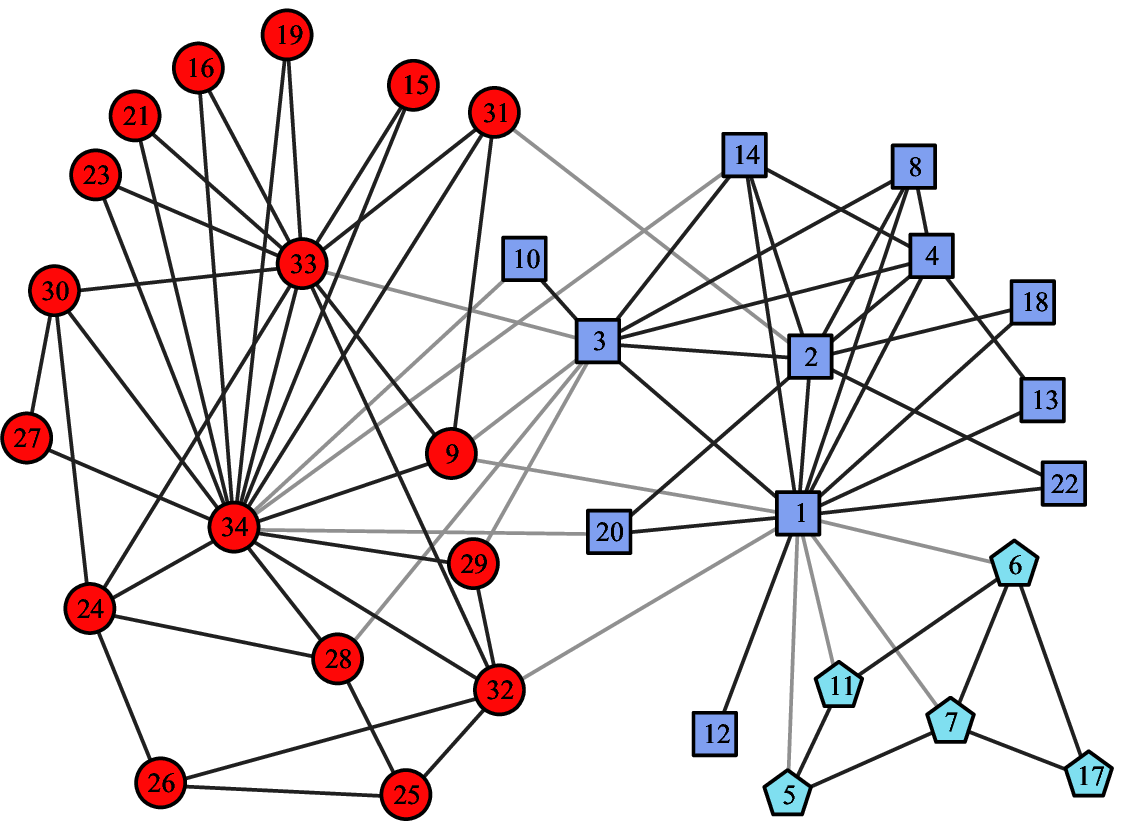}
}\\
\subfloat[\label{fig:karate:newman2013}]{
	\includegraphics[scale=\figscale]{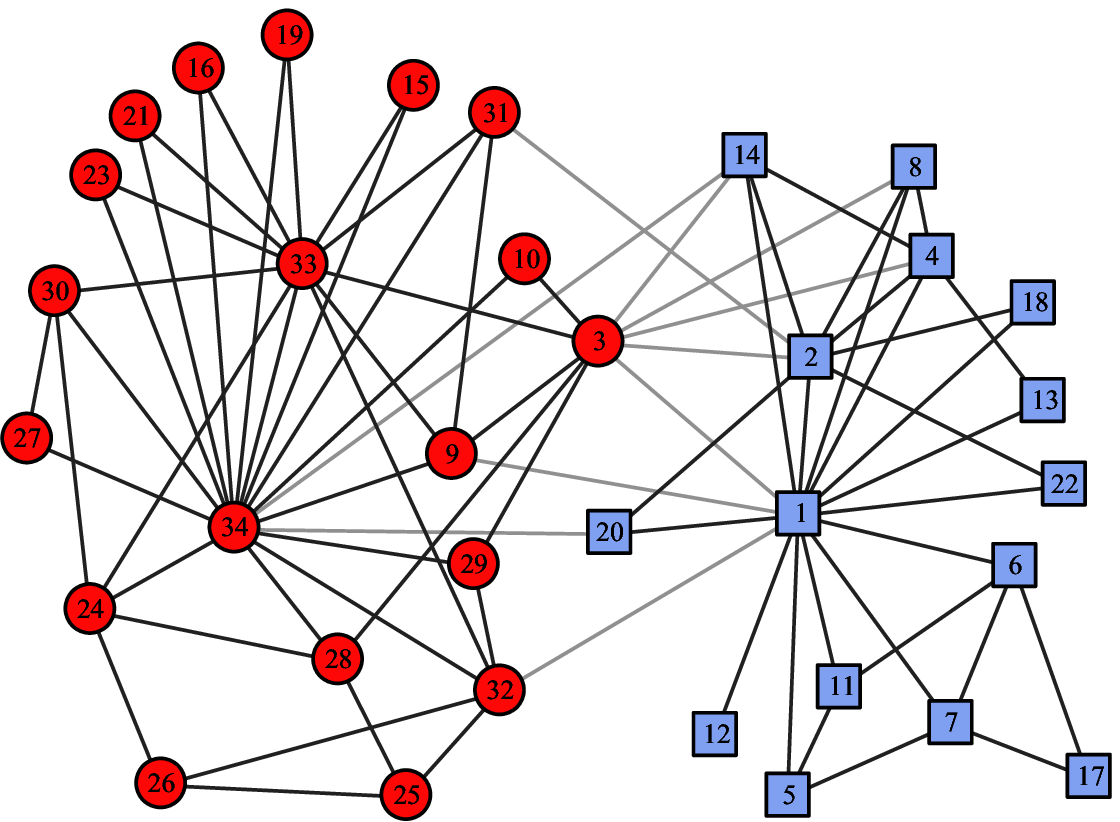}
}
\subfloat[\label{fig:karate:without_sparsification}]{
	\includegraphics[scale=\figscale]{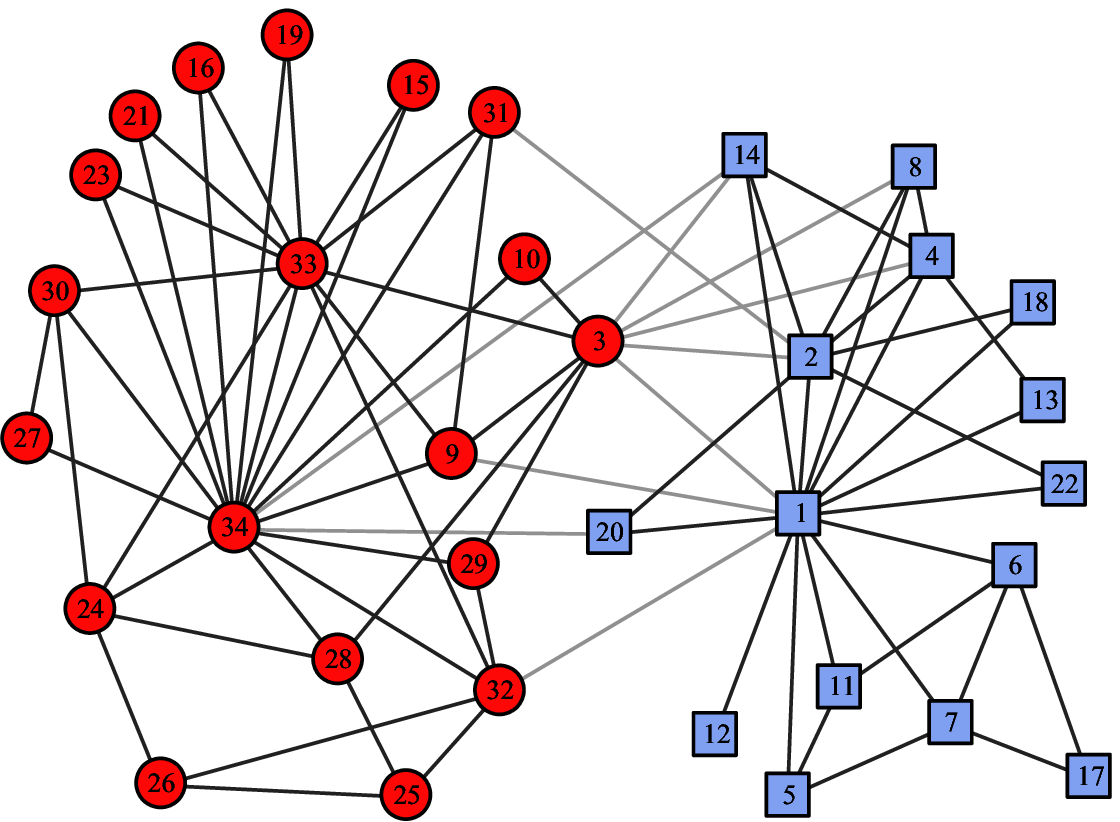}
}
\subfloat[\label{fig:karate:proposal}]{
	\includegraphics[scale=\figscale]{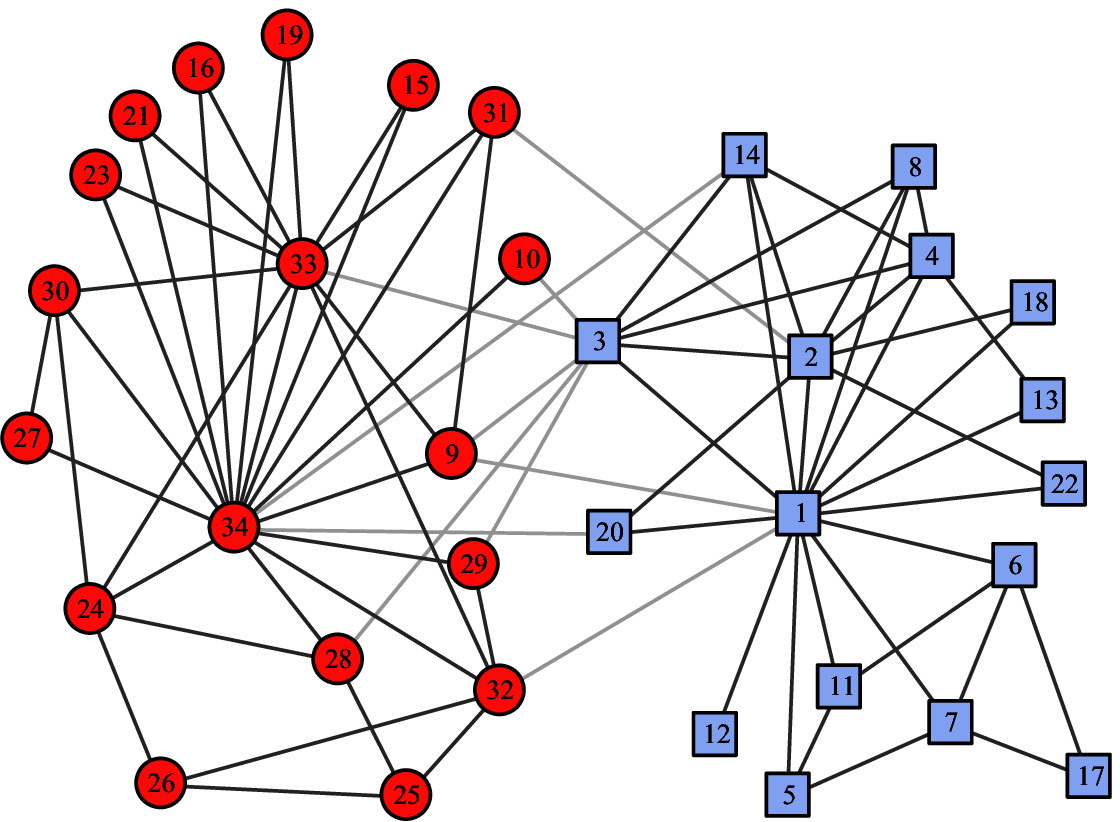}
}
\caption{\label{fig:karate}\textbf{Zachary's karate club network.} \protect\subref{fig:karate:ground_truth} The ground-truth community structure; \protect\subref{fig:karate:spectral_clustering} The community structure identified by the spectral clustering algorithm; \protect\subref{fig:karate:newman2006} The community structure detected by Newman2006; \protect\subref{fig:karate:infohiermap} The community structure revealed by Infohiermap; \protect\subref{fig:karate:newman2013} The community structure found by Newman2013; \protect\subref{fig:karate:without_sparsification} The community structure extracted by the lite version of our proposal; \protect\subref{fig:karate:proposal} The community structure extracted by the complete version of our proposed method.}
\end{figure*}

\textbf{Zachary's karate club network.} This network contains 34 vertices and 78 edges, in which vertices represent members of a karate club, edges represent social interactions between members being observed within or away from the karate club. Later, the club split into two factions because of a dispute between the administrator and the instructor. Matched with the two factions, the network contains two communities, whose structure is shown in Figure \ref{fig:karate:ground_truth}. Feeding this network into the comparison algorithms and our proposed method, we obtained the results shown in Figures \ref{fig:karate:spectral_clustering}--\ref{fig:karate:proposal}, respectively. And the values of the three metrics obtained on this network are listed in Table \ref{tab:comparisons}.

\newcommand{\tabincell}[2]{\begin{tabular}{@{}#1@{}}#2\end{tabular}}

On this network, in the result of the spectral clustering algorithm, one vertex is classified in the incorrect community. For Newman2006, although it is originated from modularity optimization, the modularity of its result is smaller than that of Infohiermap, the latter is the highest on this network, but both of their community structures deviate far from the ground truth. Newman2013 bisected the network into two communities with one vertex being misclassified also. For the lite version of our proposal, it obtained the same result as Newman2013, which is not a coincidence because the matrix under them are both derived from Eq. \ref{eq:generalized_eigenvector_equation}. Compared with them, The result extracted by the complete version of our proposed method is identical to the ground-truth community structure, i.e., our proposal acquired the best result on this network.

\begin{figure*}[ht]
\centering
\includegraphics[scale=0.475]{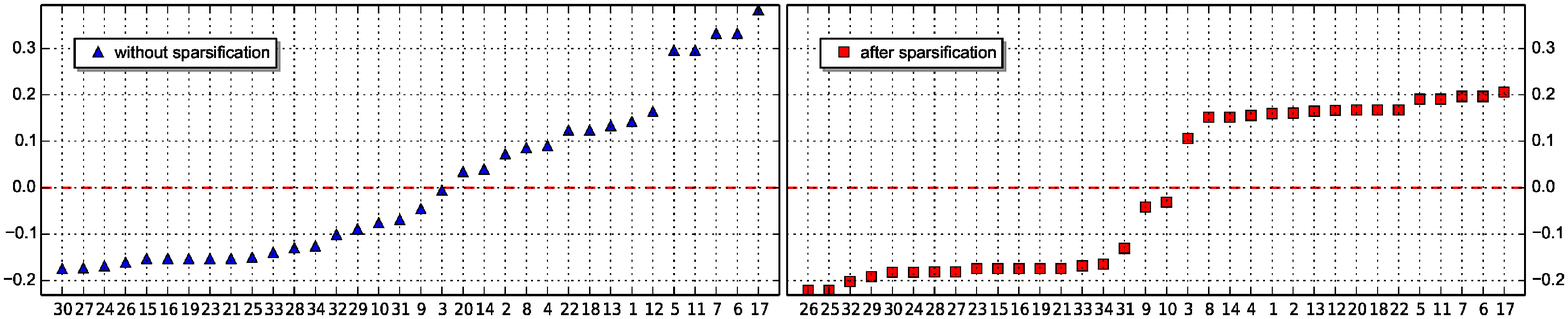}
\caption{\label{fig:karate:eigenvector_elements}\textbf{The change of values of the second eigenvector elements on Zachary's karate club network.} The left panel shows the case of the original network without sparsification, and the right panel is the case corresponding to the network after sparsification.}
\end{figure*}

Furthermore, the change of values of the second eigenvector elements on this network without sparsification and after sparsification is illustrated in Figure \ref{fig:karate:eigenvector_elements}. The gap between the positive elements and the negative elements of the second eigenvector after sparsification is much larger than the counterpart without sparsification apparently, which means that the boundary between the two communities becomes more clearer and more sharper because of the sparsification, which demonstrates the effectiveness of the proposed network sparsification algorithm to some extent.

\textbf{Lusseau's bottlenose dolphin social network.} This network consists of 62 vertices and 159 edges, in which vertices represent bottlenose dolphins living in Doubtful Sound, New Zealand. If two dolphins are observed to be co-occurring more often than expected occasionally, there is an edge between them representing their association.

\begin{figure*}
\centering
\newcommand\figscale{0.46}
\subfloat[\label{fig:dolphins:ground_truth}]{
  \includegraphics[scale=\figscale]{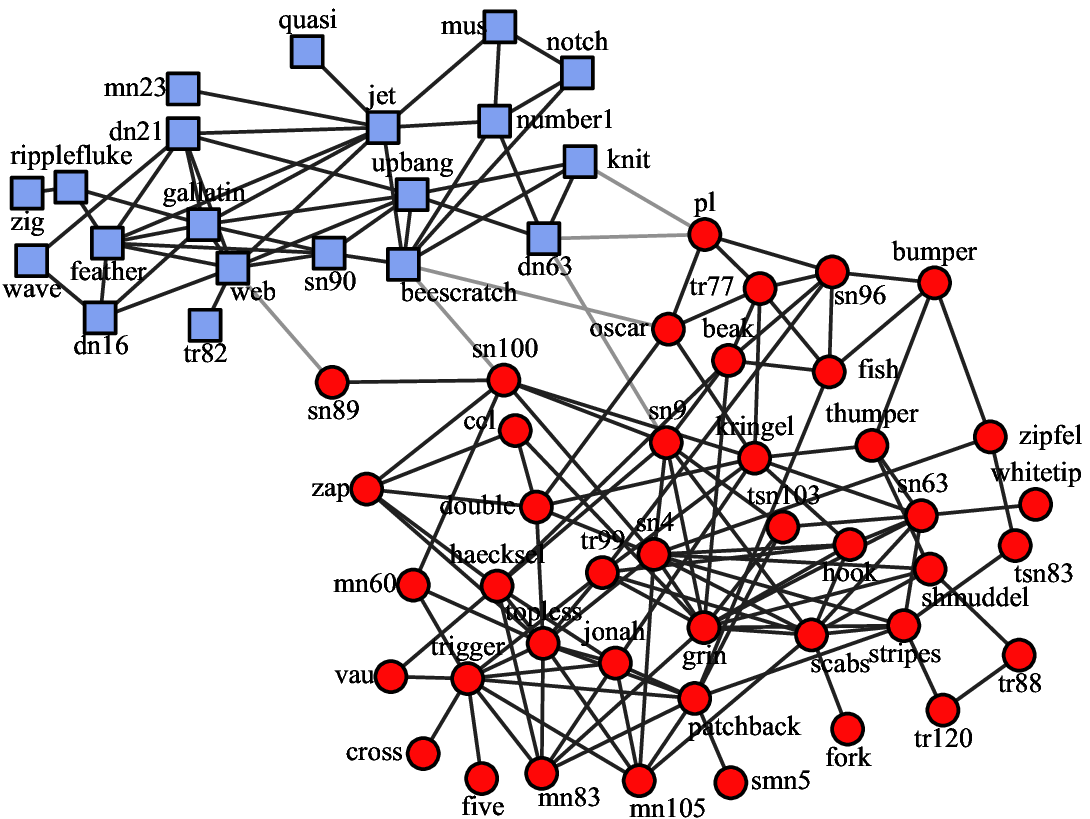}	
}\\
\subfloat[\label{fig:dolphins:spectral_clustering}]{
  \includegraphics[scale=\figscale]{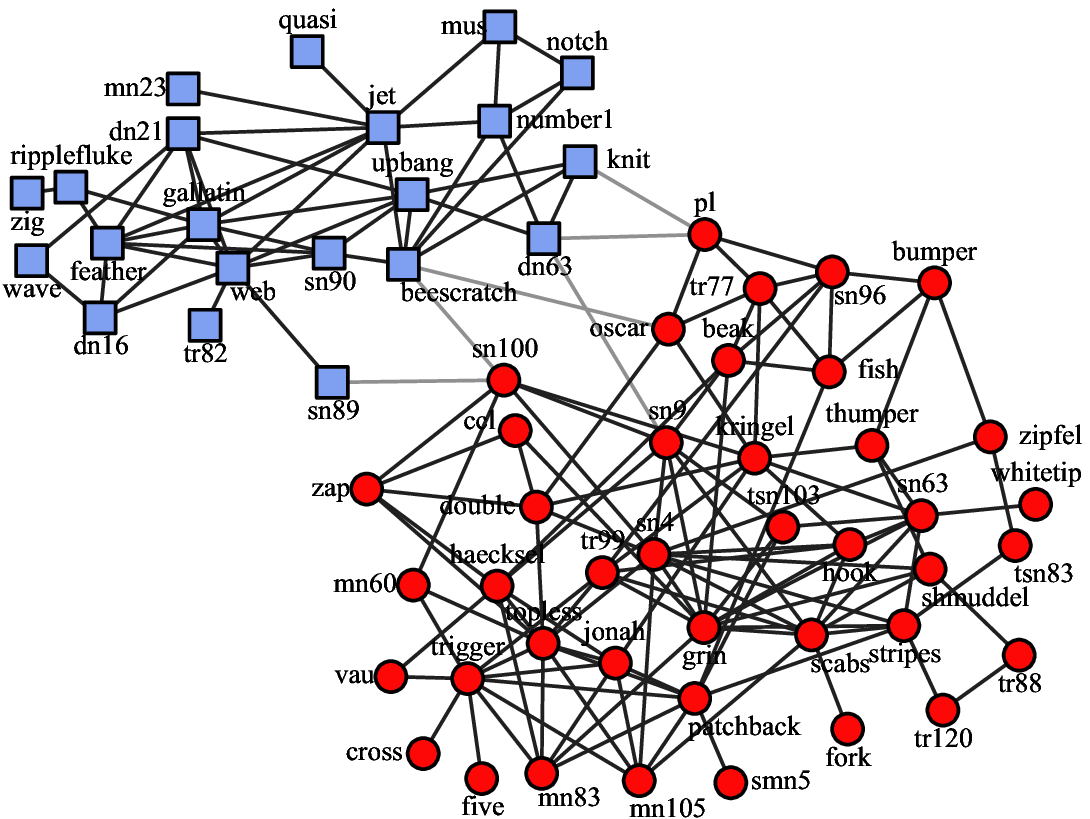}
}%
\subfloat[\label{fig:dolphins:newman2006}]{
	\includegraphics[scale=\figscale]{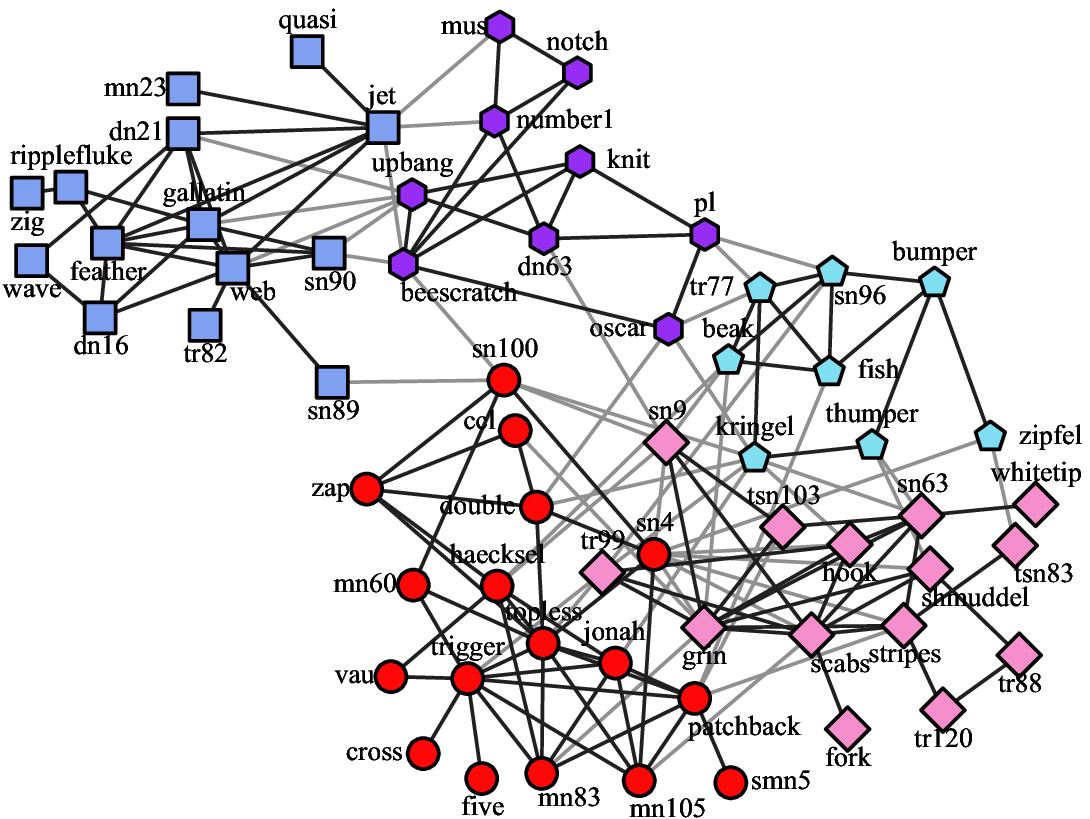}
}%
\subfloat[\label{fig:dolphins:infohiermap}]{
	\includegraphics[scale=\figscale]{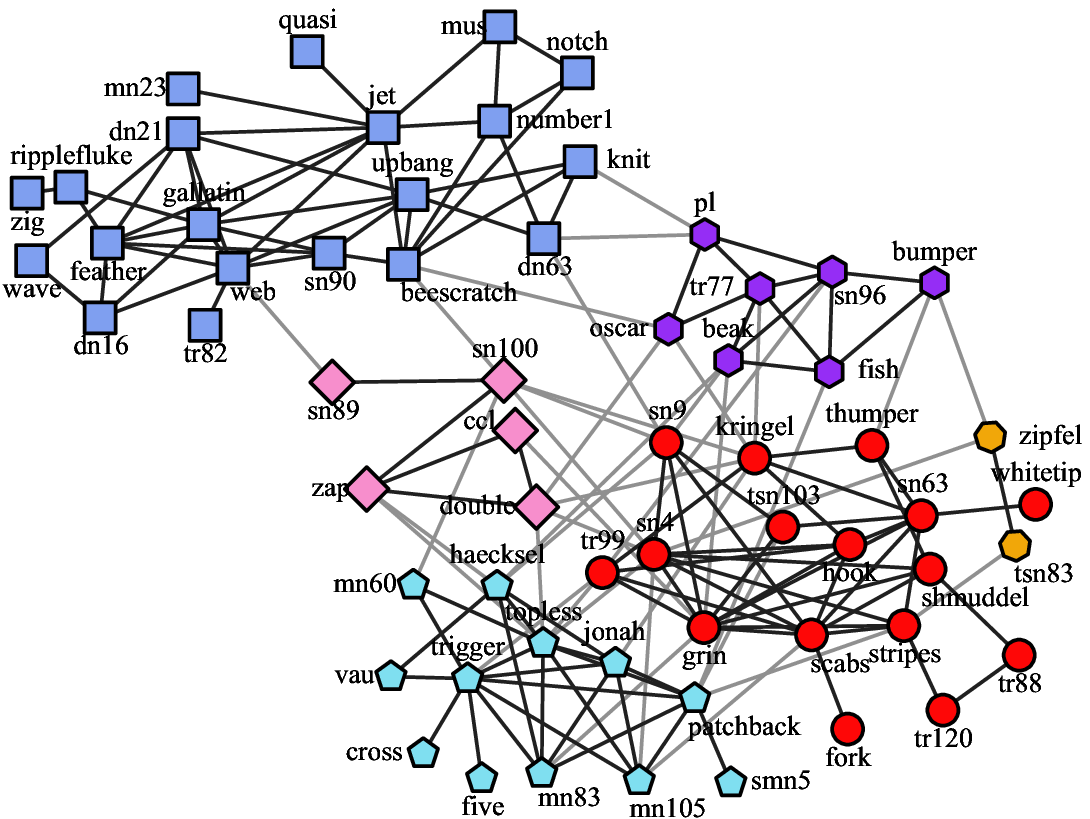}
}\\
\subfloat[\label{fig:dolphins:newman2013}]{
	\includegraphics[scale=\figscale]{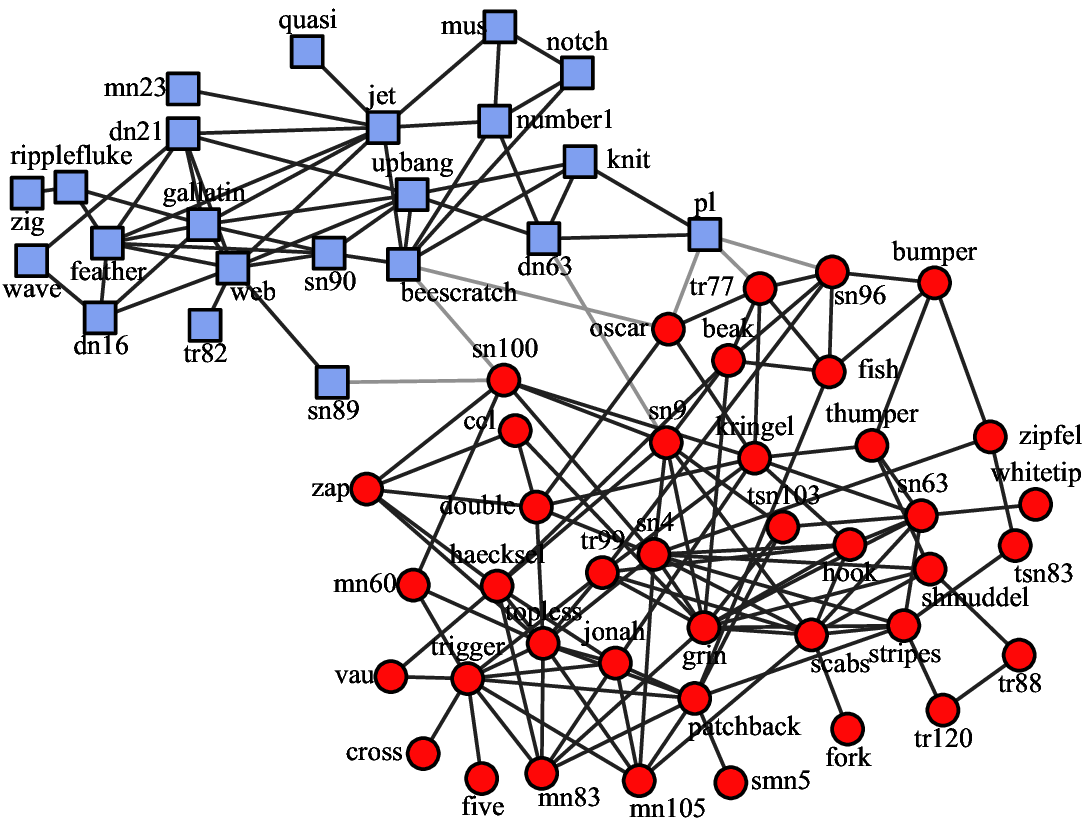}
}%
\subfloat[\label{fig:dolphins:without_sparsification}]{
	\includegraphics[scale=\figscale]{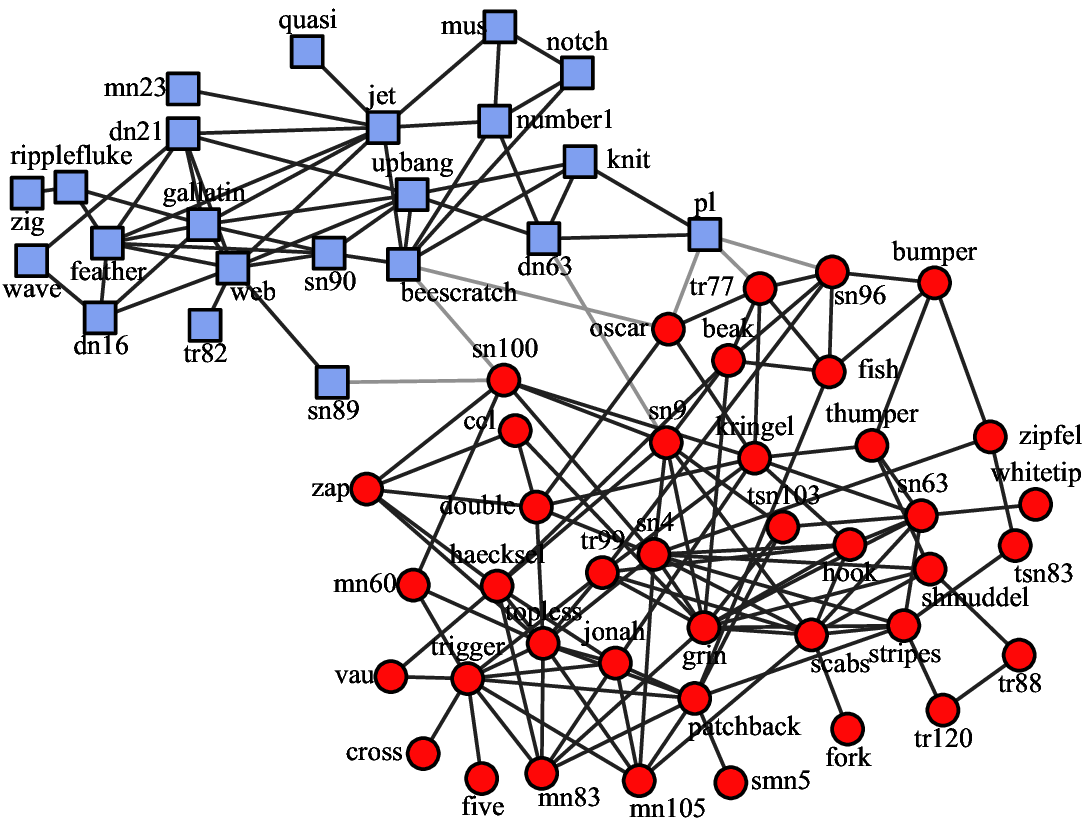}
}%
\subfloat[\label{fig:dolphins:proposal}]{
	\includegraphics[scale=\figscale]{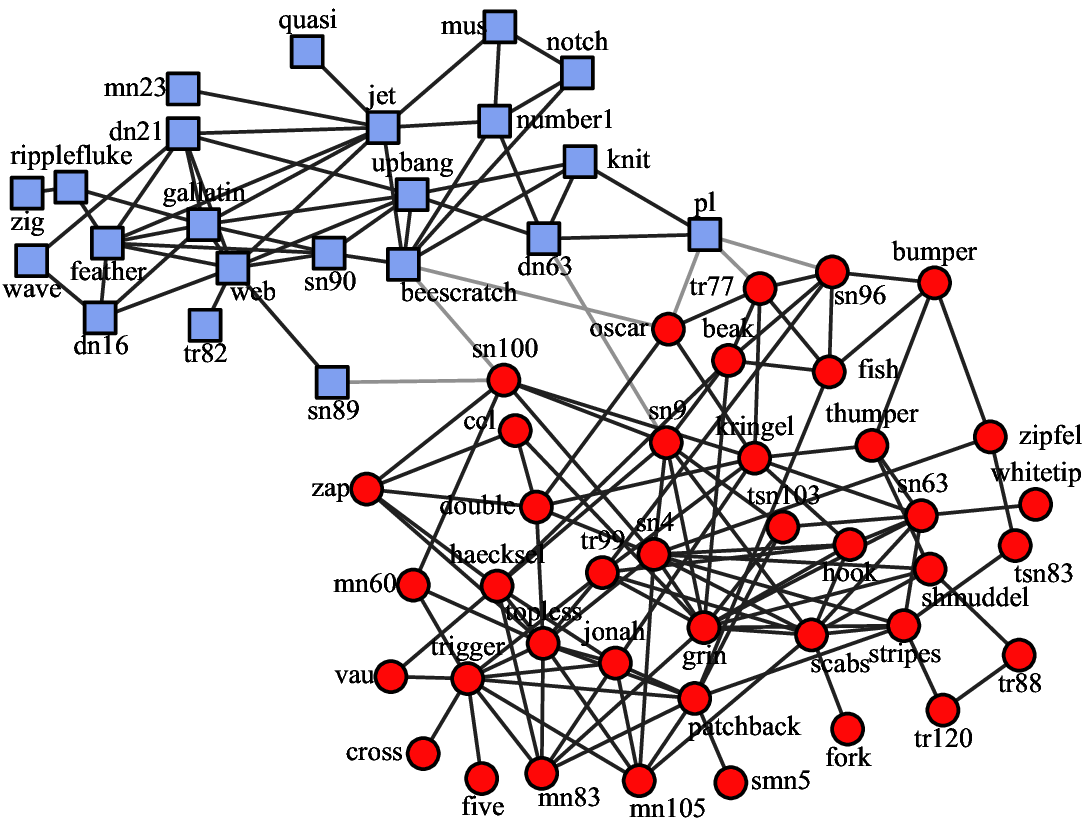}
}

\caption{\label{fig:dolphins}\textbf{Lusseau's bottlenose dolphin social network.} \protect\subref{fig:dolphins:ground_truth} The ground-truth community structure; \protect \subref{fig:dolphins:spectral_clustering} The community structure detected by the spectral clustering algorithm; \protect\subref{fig:dolphins:newman2006} The community structure extracted by Newman2006; \protect\subref{fig:dolphins:infohiermap} The community structure identified by Infohiermap; \protect\subref{fig:dolphins:newman2013} The community structure revealed by Newman2013; \protect\subref{fig:dolphins:without_sparsification} The community structure uncovered by the lite version of the proposal; \protect\subref{fig:dolphins:proposal} The community structure detected by the complete version of our proposal.}

\vskip2ex
\includegraphics[scale=0.475]{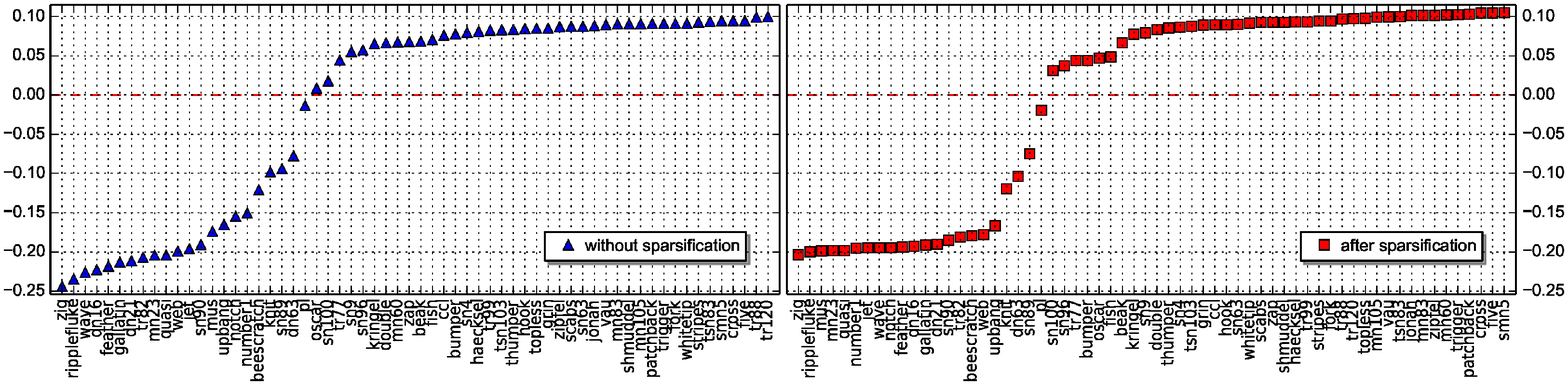}
\caption{\label{fig:dolphins:eigvec}\textbf{The change of values of the second eigenvector elements on Lusseau's bottlenose dolphin social network.} The left panel shows the case of the original network without sparsification, and the right panel presents the case after sparsification.}

\end{figure*}

The ground-truth community structure of this network is illustrated in Figure \ref{fig:dolphins:ground_truth}\footnote{Although, this network can also be considered containing 4 communities, we take it as a two-community network in this paper as in Ref. \cite{Newman:Phys.Rev.E:2013}.}, Figures \ref{fig:dolphins:spectral_clustering}--\ref{fig:dolphins:proposal} show the resulting community structures extracted by the comparison algorithms and the proposed method, individually, and the values of the three metrics acquired on this network are also filled in Table \ref{tab:comparisons}. 

On this network, the spectral clustering algorithm got the result closest to the ground-truth community structure, in which only one vertex was misclassified. For Newman2006 and Infohiermap, they both extracted more than two communities from this network, but both of them differ far from the ground truth\footnote{In addition, they also departure far from the four-community ground-truth structure of this network.}. For Newman2013, for the lite version and the complete version of our proposed method, they all identified the same community structure from this network, in which two vertices were wrongly classified into the opposite community. It seems that the proposed network sparsification algorithm failed to sparsify the network, so that the complete version of our proposal did not acquire the better result than that of without sparsification. But that is not the case. To demonstrate the case, as did in Figure \ref{fig:karate:eigenvector_elements}, we also plotted the change of values of the second eigenvector elements on this network without sparsification and after sparsification in Figure \ref{fig:dolphins:eigvec}. Evidently, the gap between the positive elements and the negative elements is also much larger after sparsification than that without sparsification, which means that our proposed network sparsification algorithm does take its effect on this network. %

\begin{figure*}[!htbp]
  \centering
  \newcommand{\figscale}{0.45}
  \subfloat[\label{fig:riskmap:ground_truth}]{
    \includegraphics[scale=\figscale]{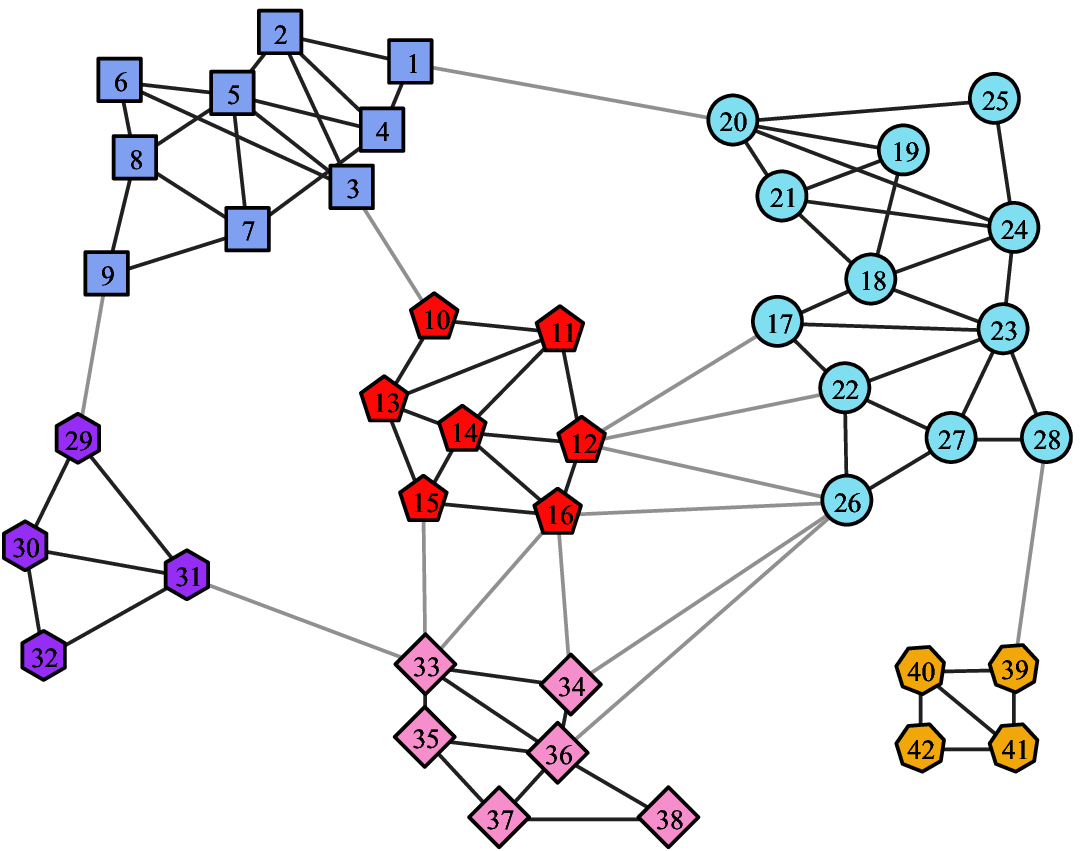}
  }
  \subfloat[\label{fig:riskmap:spectral_clustering}]{
    \includegraphics[scale=\figscale]{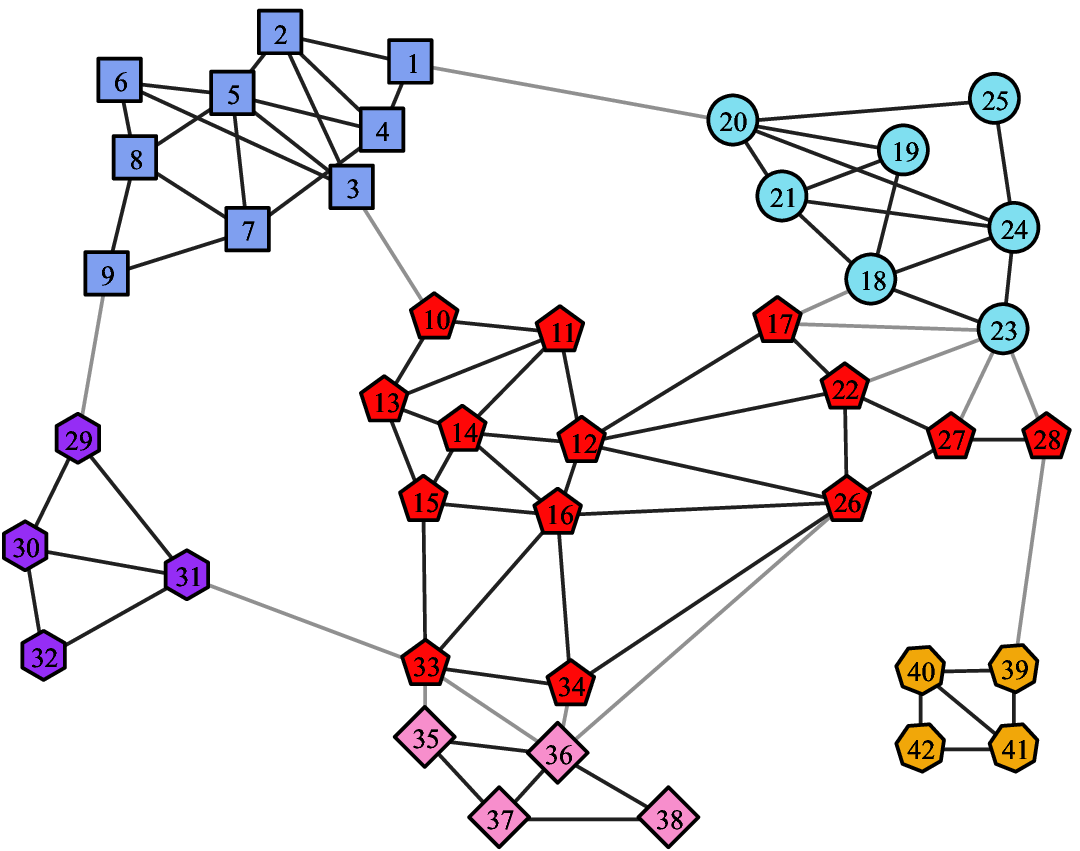}
  }
  \subfloat[\label{fig:riskmap:newman2006}]{
    \includegraphics[scale=\figscale]{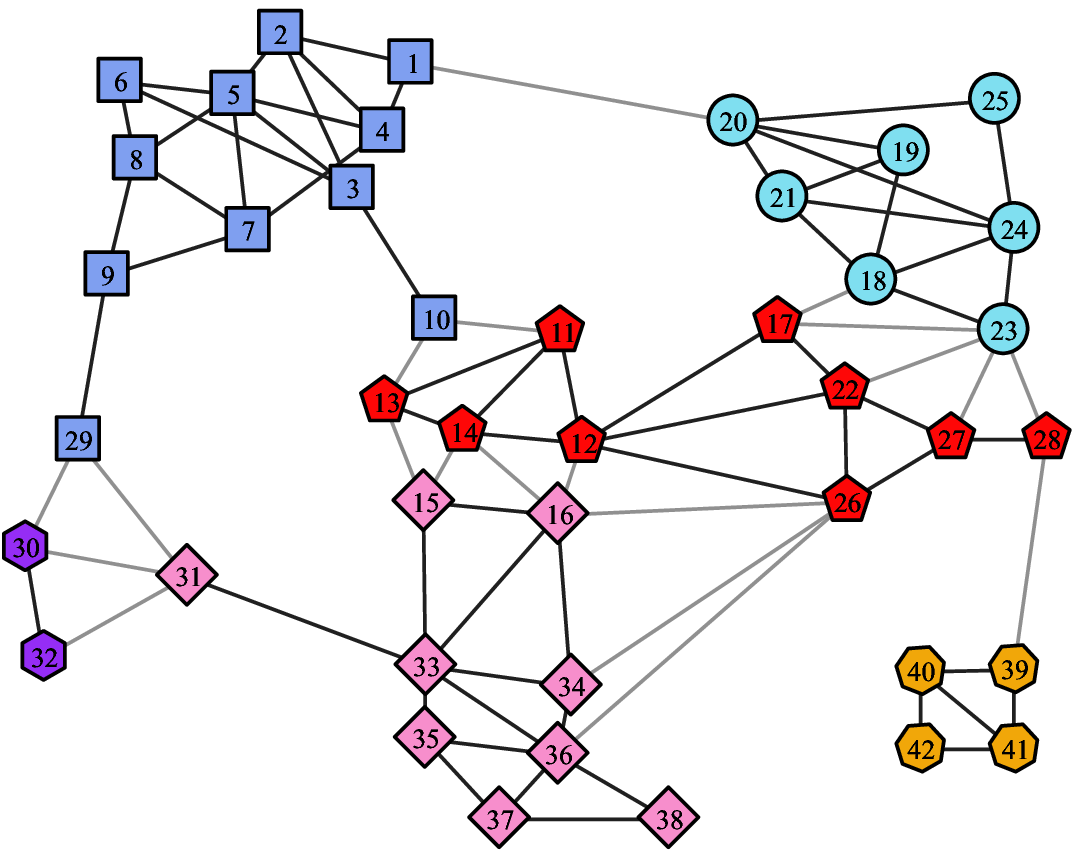}
  }\\
  \subfloat[\label{fig:riskmap:infohiermap}]{
    \includegraphics[scale=\figscale]{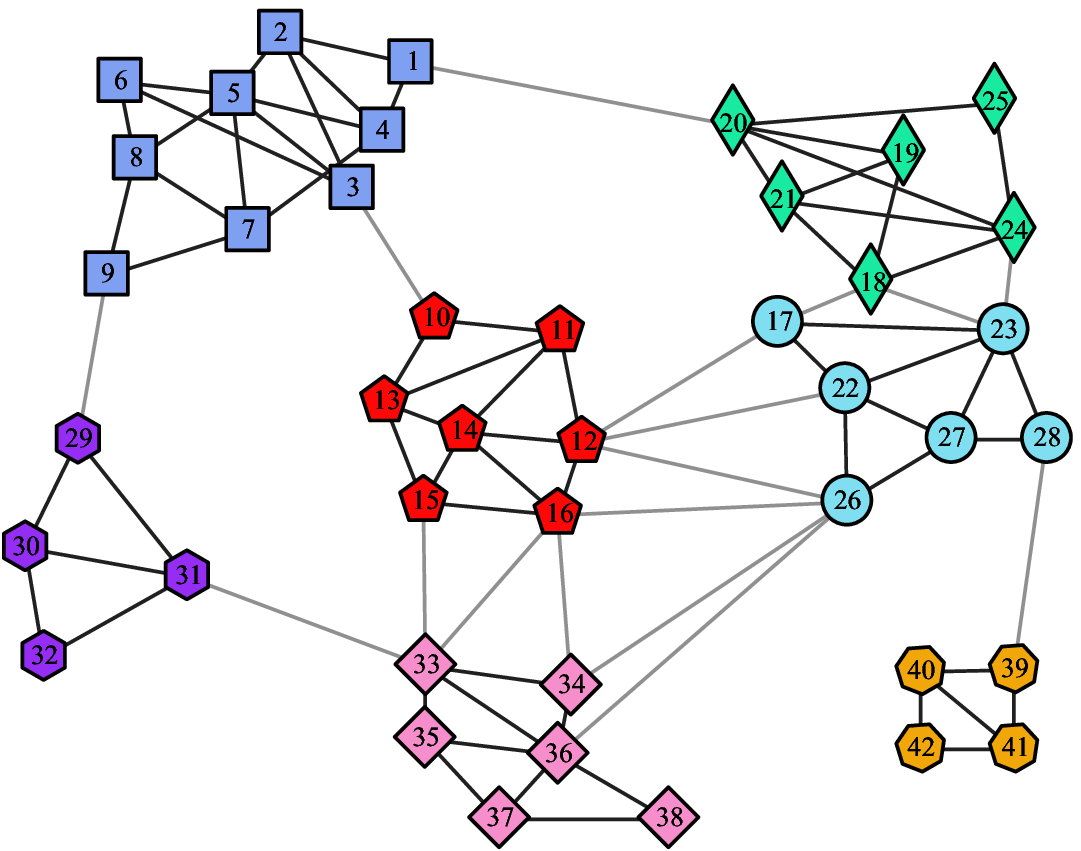}
  }
  \subfloat[\label{fig:riskmap:without_sparsification}]{
    \includegraphics[scale=\figscale]{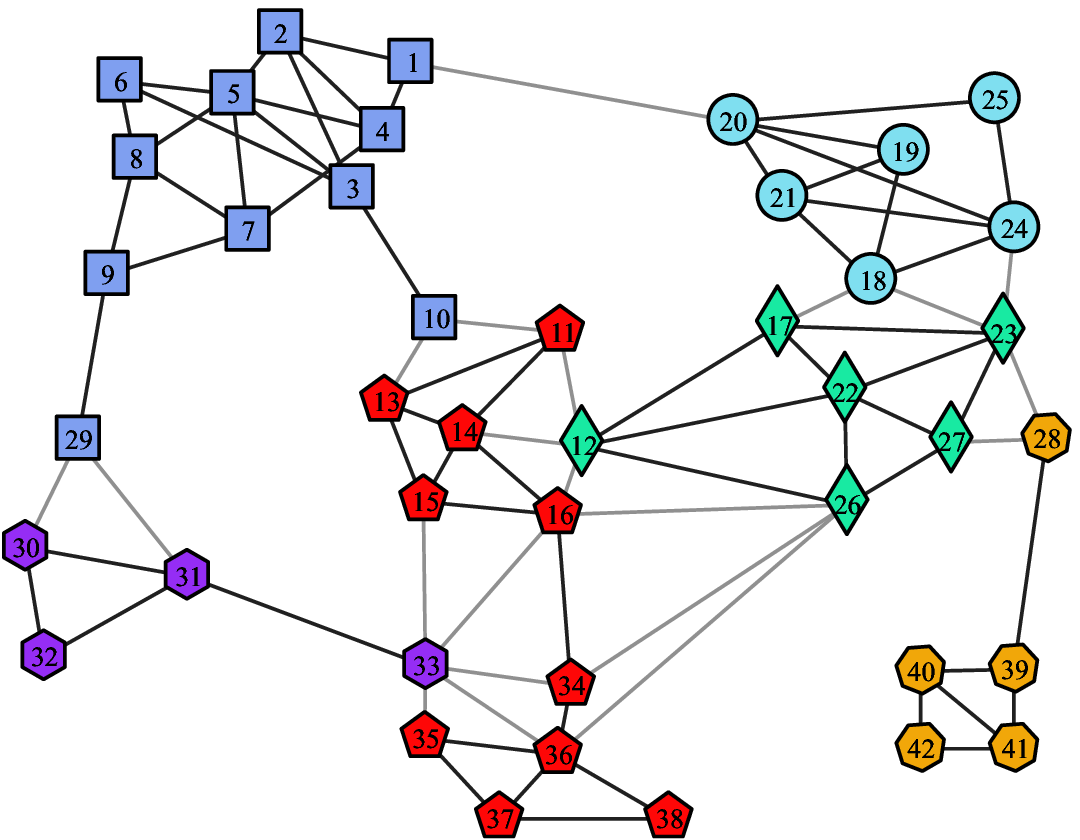}
  }
  \subfloat[\label{fig:riskmap:proposal}]{
    \includegraphics[scale=\figscale]{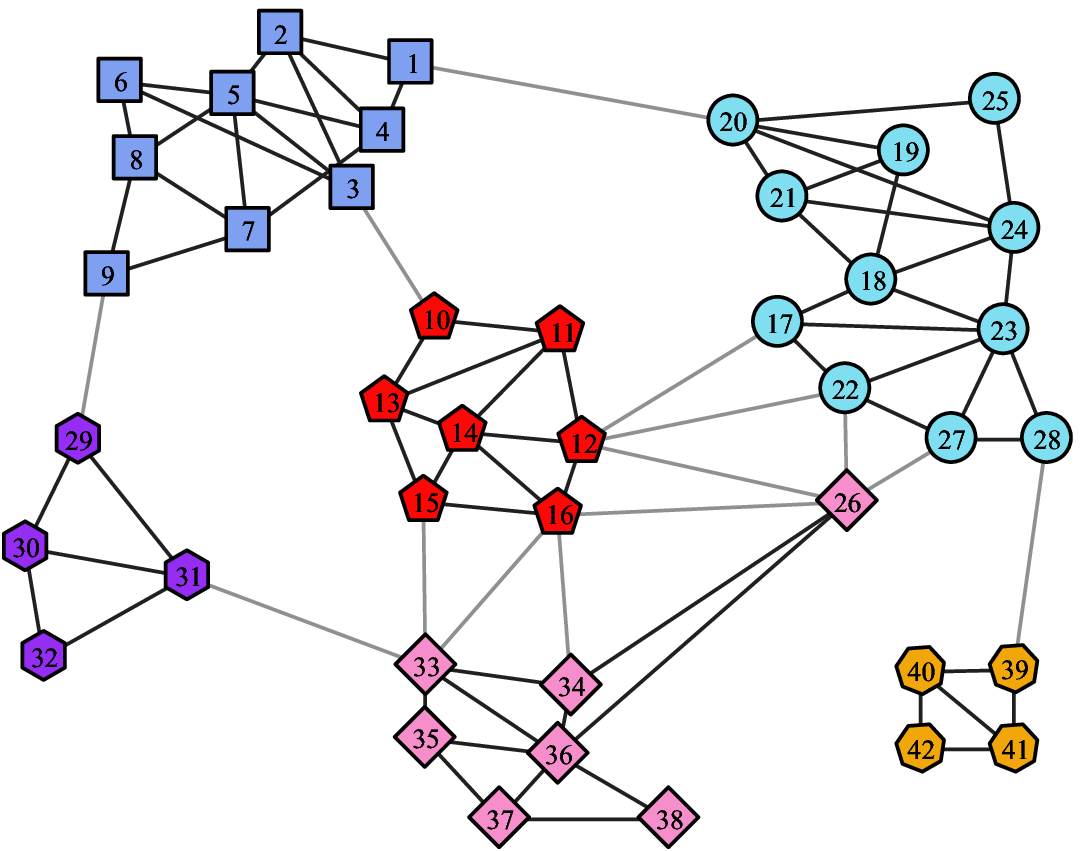}
  }
  \caption{\label{fig:riskmap}\textbf{Risk map network.} \protect\subref{fig:riskmap:ground_truth} The ground-truth community structure; \protect\subref{fig:riskmap:spectral_clustering} The community structure detected by the spectral clustering algorithm; \protect\subref{fig:riskmap:newman2006} The community structure found by Newman2006; \protect\subref{fig:riskmap:infohiermap} The community structure revealed by Infohiermap; \protect\subref{fig:riskmap:without_sparsification} The community structure identified by the lite version of the proposal; \protect\subref{fig:riskmap:proposal} The community structure detected by the complete version of our proposed method.}
\end{figure*}

\textbf{Risk map network.} This network is a map of the popular strategy board game, Risk\footnote{\url{https://en.wikipedia.org/?title=Risk_(game)}}. It is a political map of the Earth, divided into 42 territories, which are grouped into 6 continents. Therefore, this network is comprised of 42 vertices and 83 edges. In accordance with the 6 continents naturally, the ground-truth community structure of this network is as shown in Figure \ref{fig:riskmap:ground_truth}, running the comparison algorithms and the proposed method on this network, we obtained the results illustrated in Figures \ref{fig:riskmap:spectral_clustering}--\ref{fig:riskmap:proposal}, respectively, and the values of the three metrics achieved on this network are enumerated in Table \ref{tab:comparisons} as well.

In this network, vertices ``26'', ``12'', ``16'', and ``33'' are special ones. Taking vertex ``26'' as an example, there are 6 edges associated with it, but they are incident to 3 different communities with 2 edges each community. It is hard to determine in which community this vertex should belong according to the topological information only. The similar scenarios occur also for the other special vertices, it is reasonable that they be classified into any community incident to them, without considering the physical meaning of the vertices. For this reason, mistakes around these vertices tend to be introduced by community detection algorithms.

All of the results of the spectral clustering algorithm, of Newman2006, and of the lite version and the complete version of our proposal contain misclassifications of one or more of these special vertices. An exception is Infohiermap, it incredibly classified all these special vertices correctly. But it split the community located at the right top of the panel into two, resulting in a lower accuracy. For our proposal, after network sparsification, the method eliminated most of the mistakes and extracted the community structure with a high degree of success, all but one of the territories are grouped correctly with the other territories in their continent, the community structure is the best one among those of other algorithms.

\begin{figure*}[!htbp]
\centering
\newcommand{\figscale}{0.38}
\subfloat[\label{fig:santafe:ground_truth}]{
	\includegraphics[scale=\figscale]{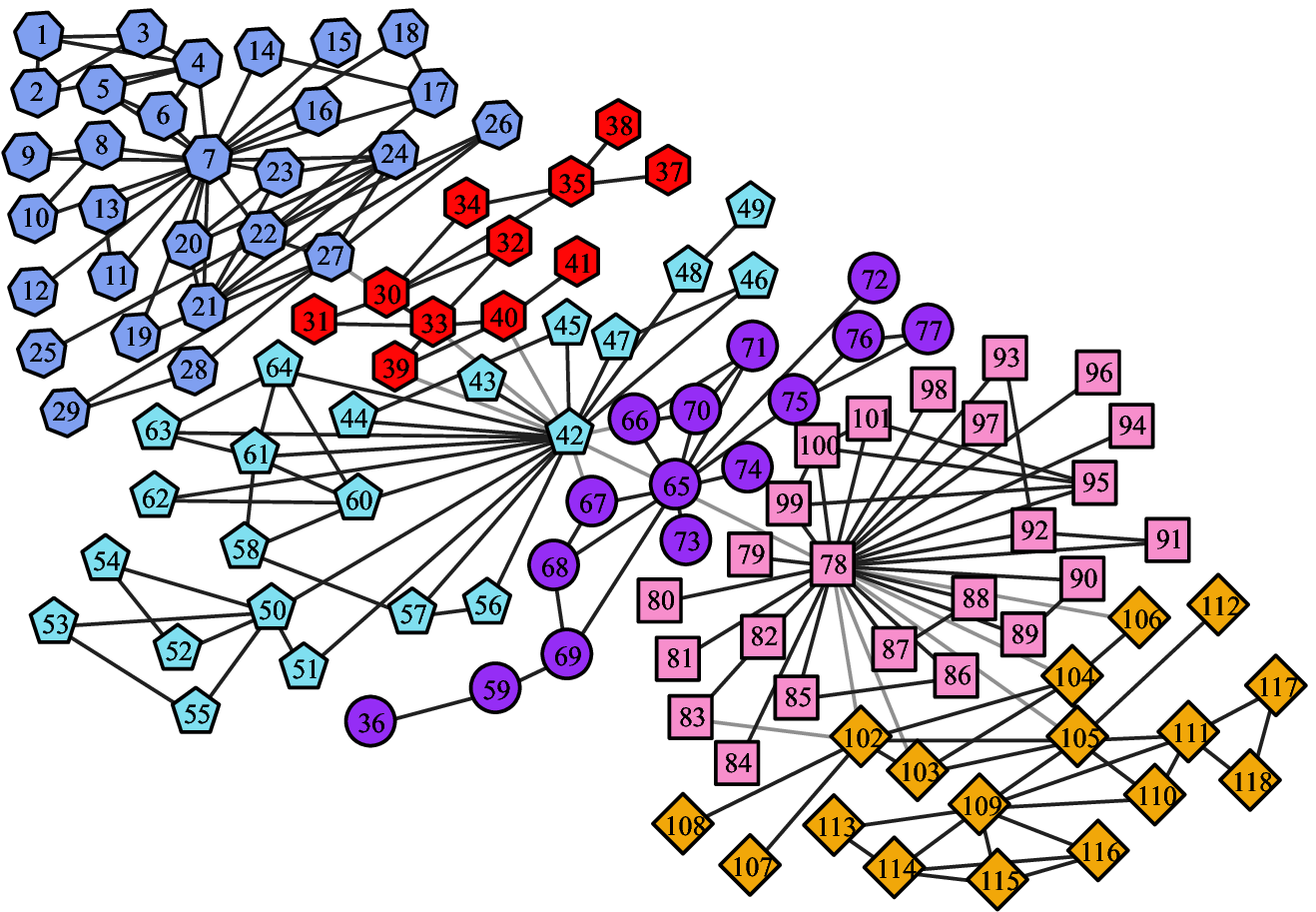}
}\\
\subfloat[\label{fig:santafe:spectral_clustering}]{
	\includegraphics[scale=\figscale]{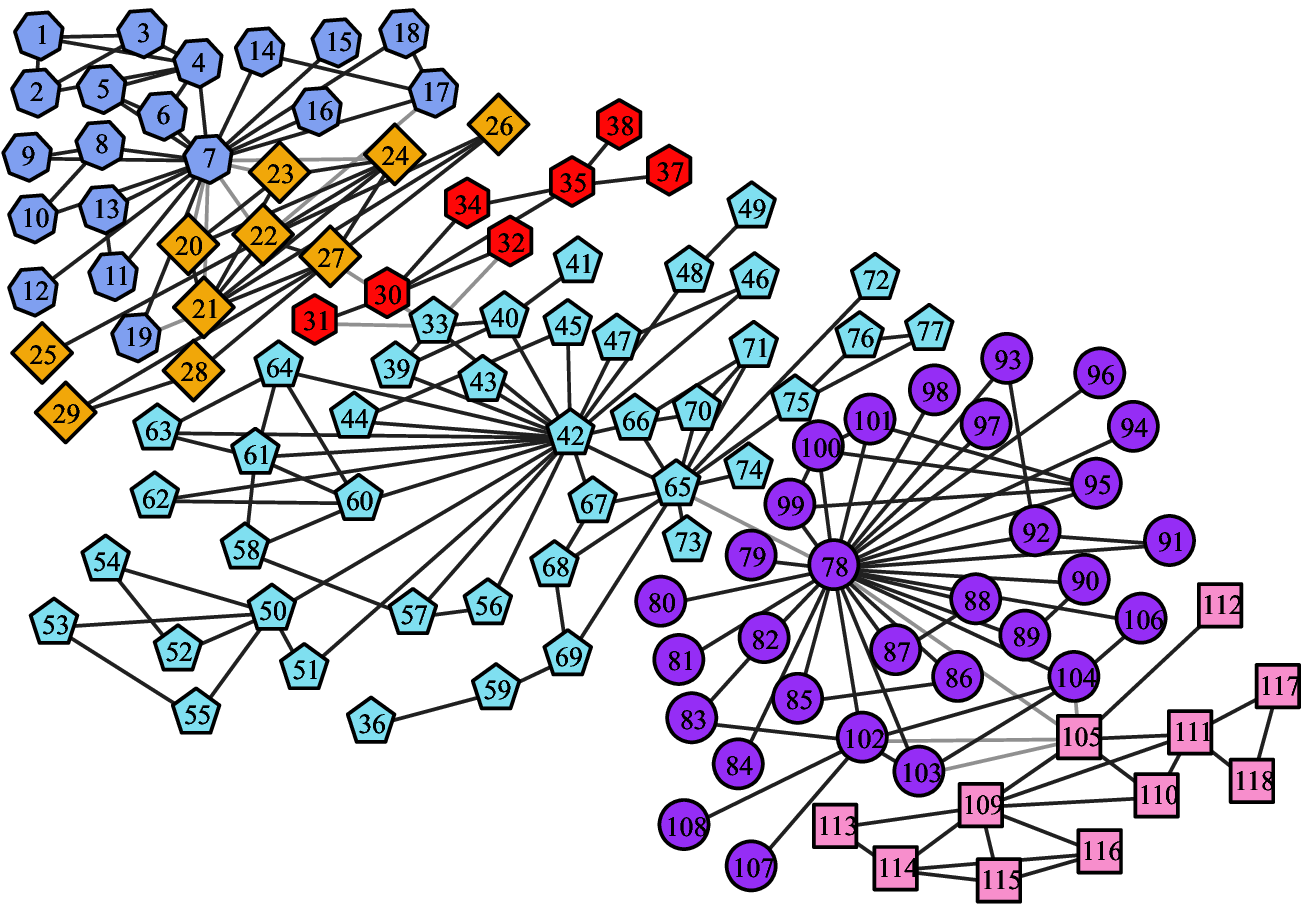}
}
\subfloat[\label{fig:santafe:newman2006}]{
	\includegraphics[scale=\figscale]{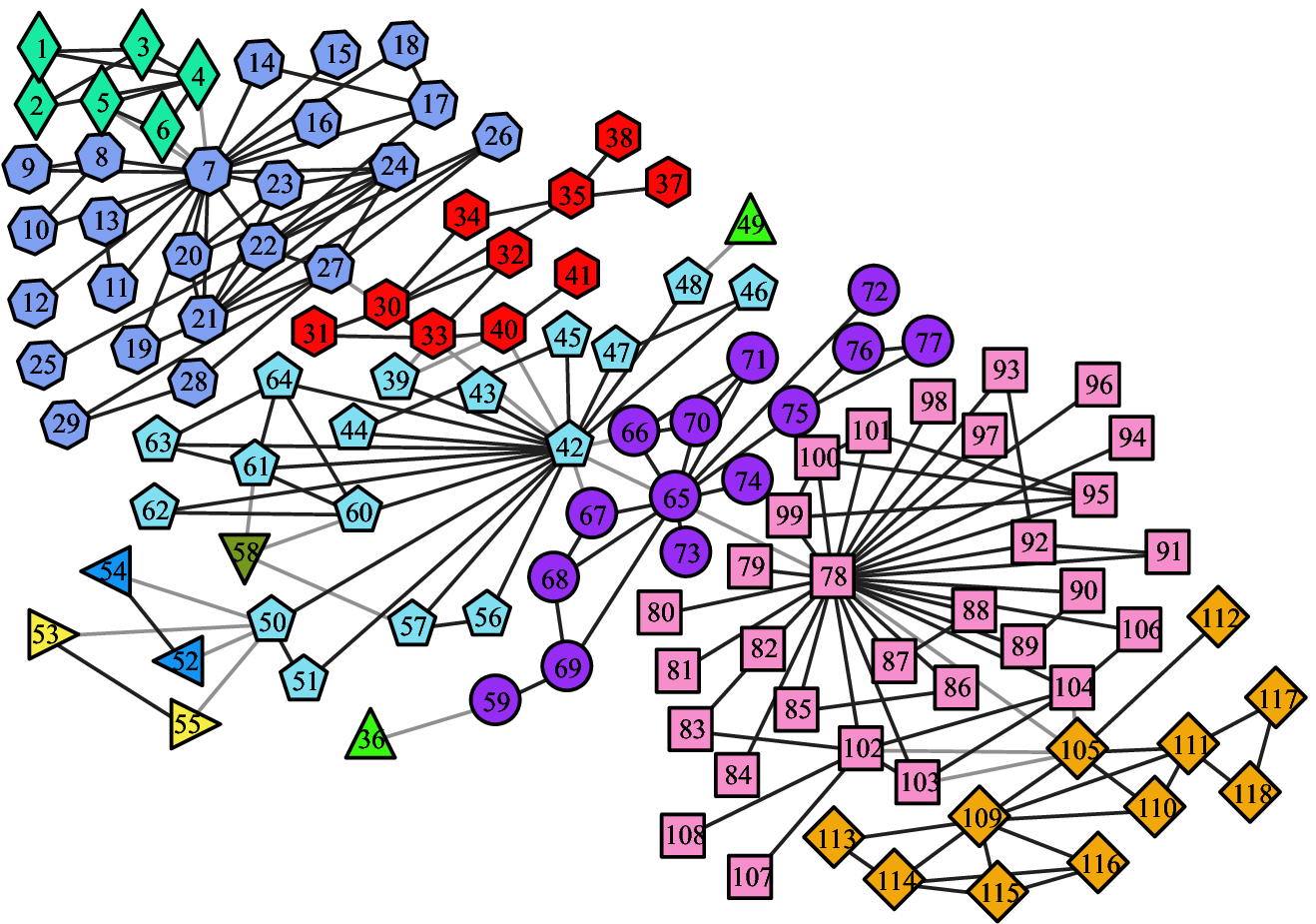}
}
\subfloat[\label{fig:santafe:infohiermap_1}]{
	\includegraphics[scale=\figscale]{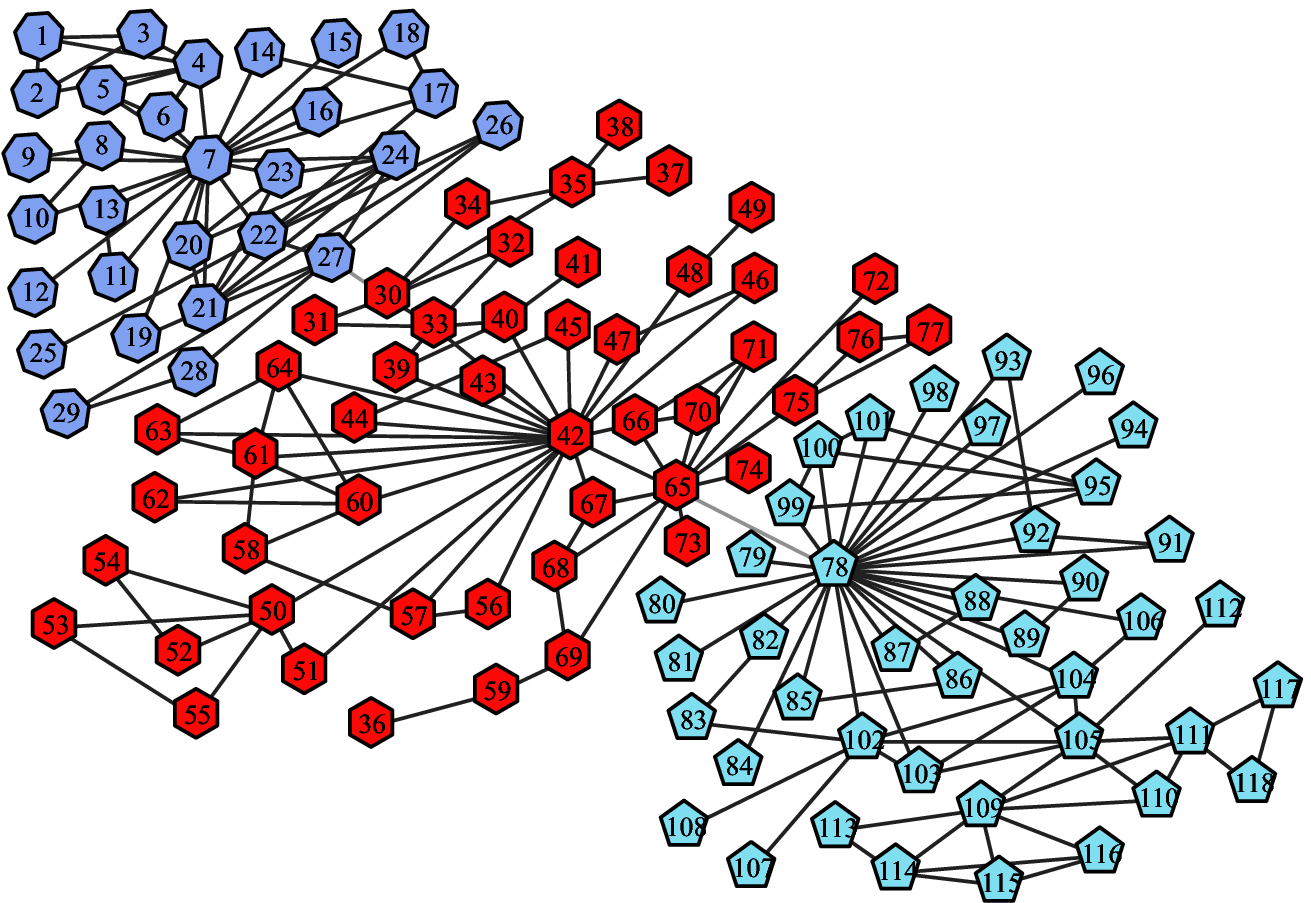}
}\\
\subfloat[\label{fig:santafe:infohiermap_2}]{
	\includegraphics[scale=\figscale]{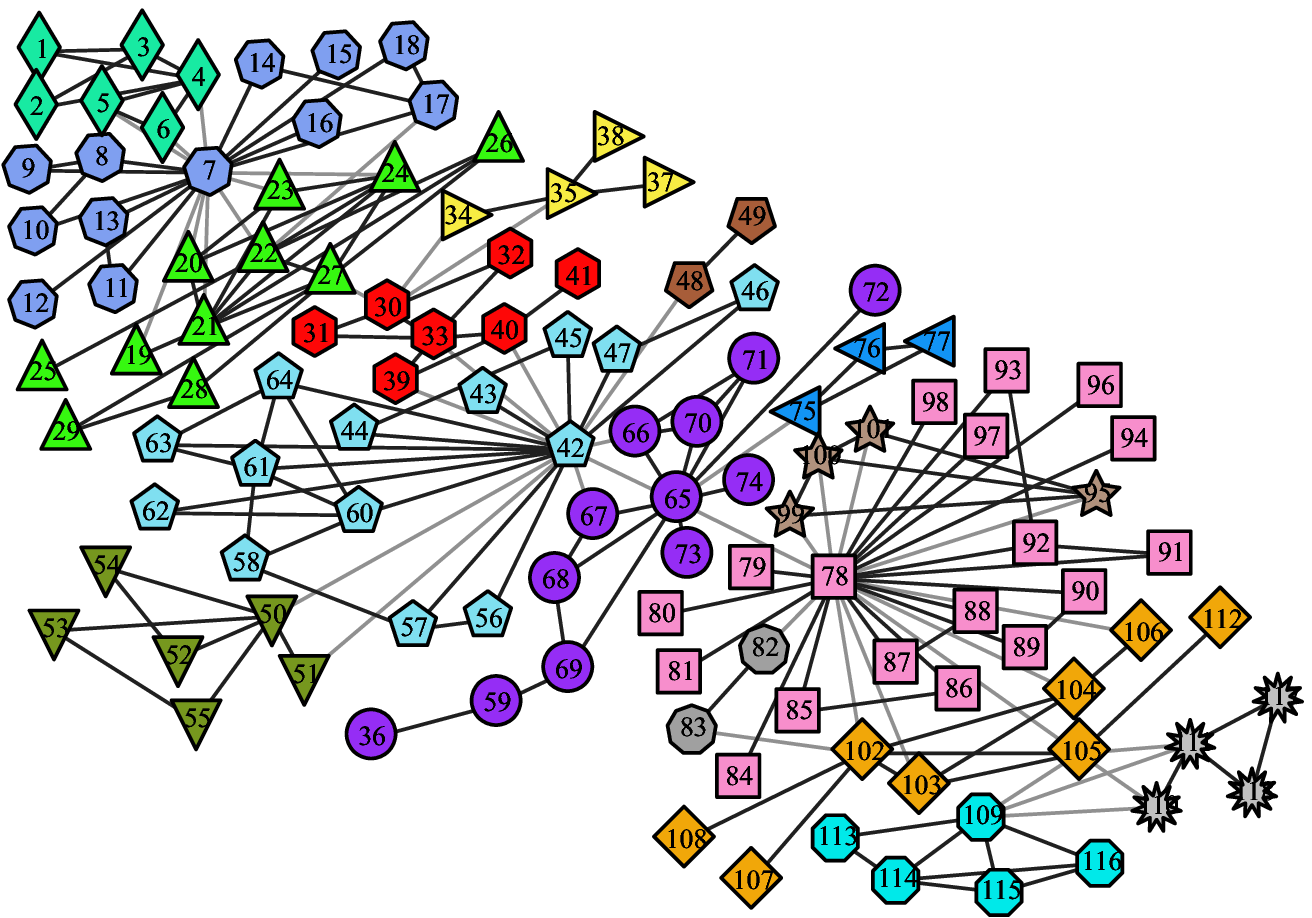}
}
\subfloat[\label{fig:santafe:without_sparsification}]{
	\includegraphics[scale=\figscale]{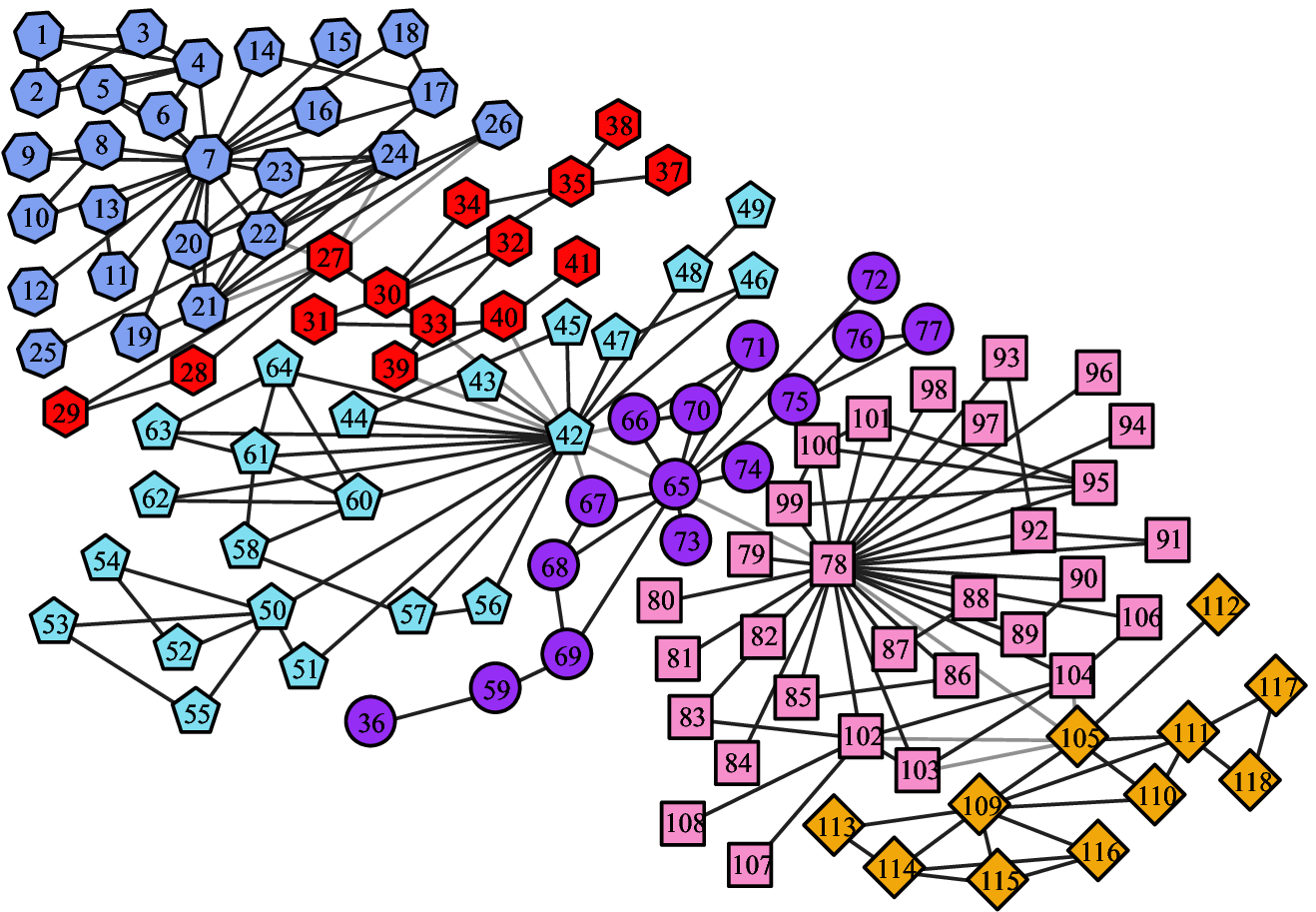}
}
\subfloat[\label{fig:santafe:proposal}]{
	\includegraphics[scale=\figscale]{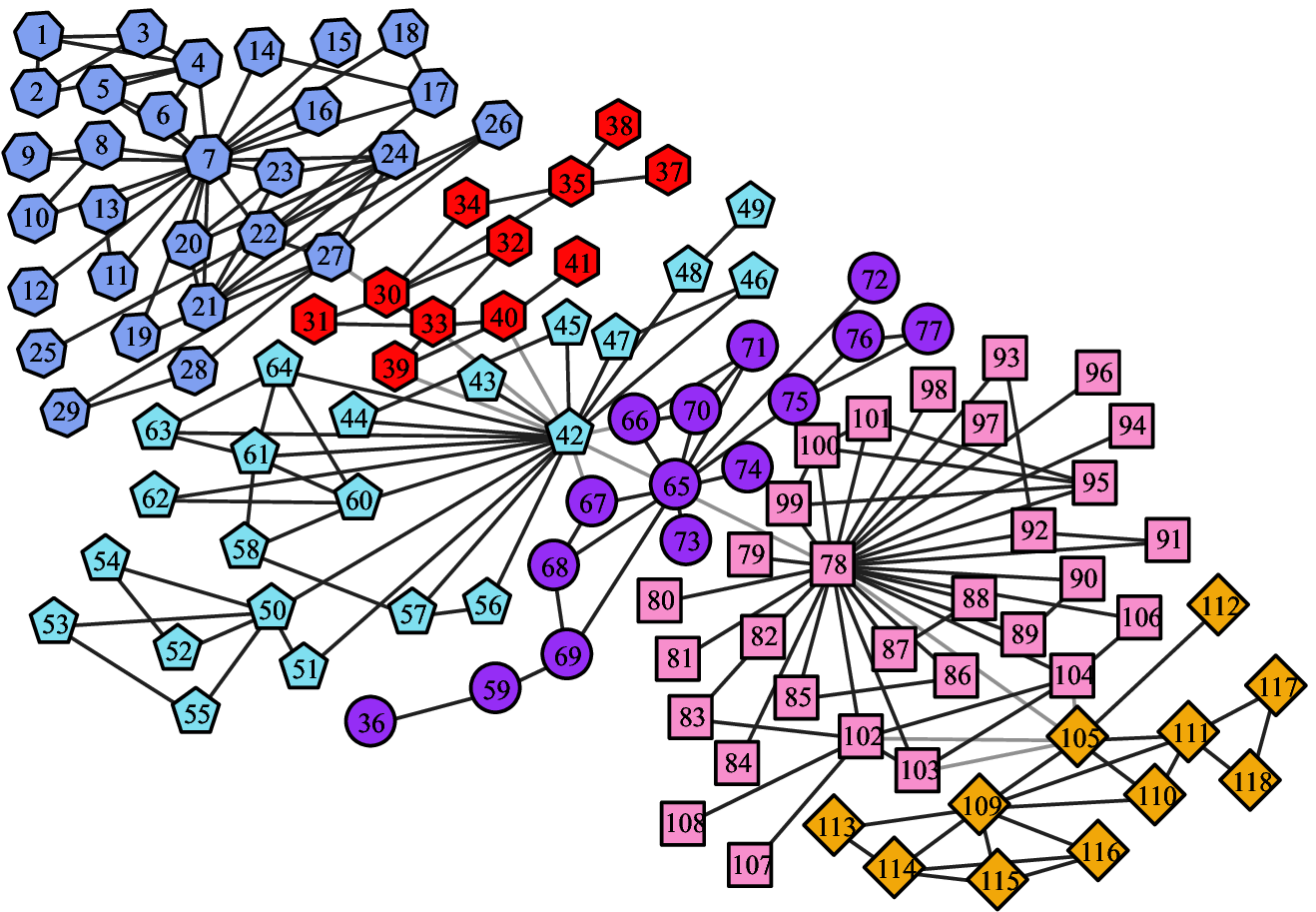}
}
\caption{\label{fig:santafe}\textbf{Scientist's collaboration network.} \protect\subref{fig:santafe:ground_truth} The ground-truth community structure; \protect\subref{fig:santafe:spectral_clustering} The community structure identified by the spectral clustering algorithm; \protect\subref{fig:santafe:newman2006} The community structure revealed by Newman2006; \protect\subref{fig:santafe:infohiermap_1} The first-level community structure extracted by Infohiermap; \protect\subref{fig:santafe:infohiermap_2} The second-level community structure extracted by Infohiermap; \protect\subref{fig:santafe:without_sparsification} The community structure found by the lite version of the proposal; \protect\subref{fig:santafe:proposal} The community structure detected by the complete version of the proposed method.}
\end{figure*}

\textbf{Scientist's collaboration network.} This network depicts coauthor relationship between 118 scientists working at the Santa Fe Institute, in which each vertex represents a scientist, and each edge connects two scientists who have coauthored at least one article. It contains 118 vertices and 197 edges, and can be naturally partitioned into 6 communities according to the scientists' specialities. The ground-truth community structure and the results extracted by the comparison algorithms and the proposed method are visualized in Figures \ref{fig:santafe:ground_truth}--\ref{fig:santafe:proposal}, severally, and the values of the three metrics are also listed in Table \ref{tab:comparisons}.

On this network, the spectral clustering algorithm merged two communities (plotted in cyan pentagon and in purple circle in Figure \ref{fig:santafe:ground_truth}, respectively) into one, but split one community (plotted in light blue heptagon in Figure \ref{fig:santafe:ground_truth}) into two. Besides this, there are 10 vertices (vertices ``33'', ``39'', ``40'', ``41'', ``102'', ``103'', ``104'', ``106'', ``107'', and ``108'') were classified into the incorrect communities, i.e., the quality of the resulting community structure is not so high. For Newman2006, the quality of the result is also quite poor, several vertex groups extracted are too trivial to be accepted as communities. For Infohiermap, it revealed two levels of community structures from this network. The first level contains only 3 communities, and the second level consists of 16 communities exaggeratively, both of them deviated far from the ground-truth community structure. In the result of the lite version of the proposed method, vertices ``27'', ``28'', ``29'', ``102'', ``103'',``104'', ``106'', ``108'', and ``109'' were misclassified, and after sparsification, the mistakes introduced on the former three vertices were eliminated by the complete version of our proposal. Unfortunately however, there are still 6 vertices that were classified into the incorrect community in the final community structure. Even though, the resulting structure of our proposed method is the one closest to the ground-truth community structure. Which means, compared with other algorithms, our proposed method extracted the best community structure from this network. 

\begin{figure*}[!htbp]
\centering
\newcommand\figscale{0.45}
\subfloat[\label{fig:football:ground_truth}]{
	\includegraphics[scale=\figscale]{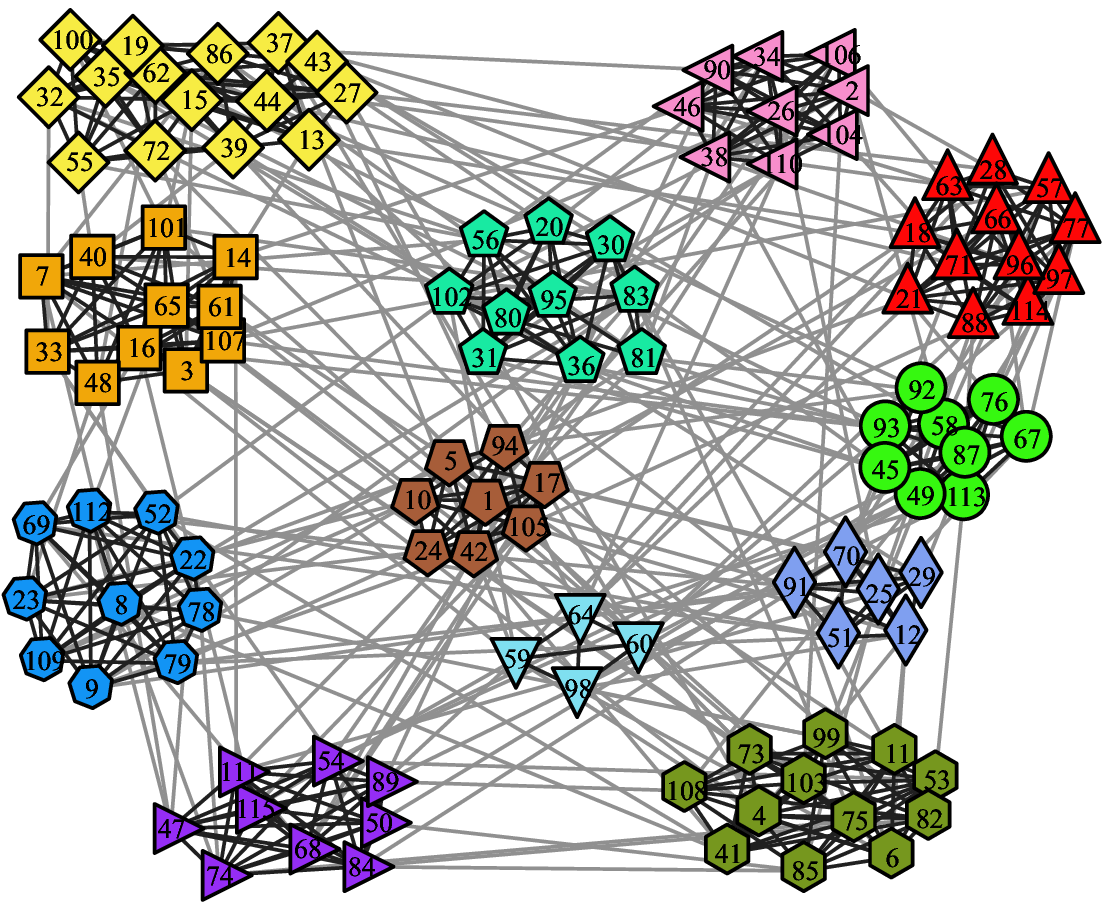}
}
\subfloat[\label{fig:football:spectral_clustering}]{
	\includegraphics[scale=\figscale]{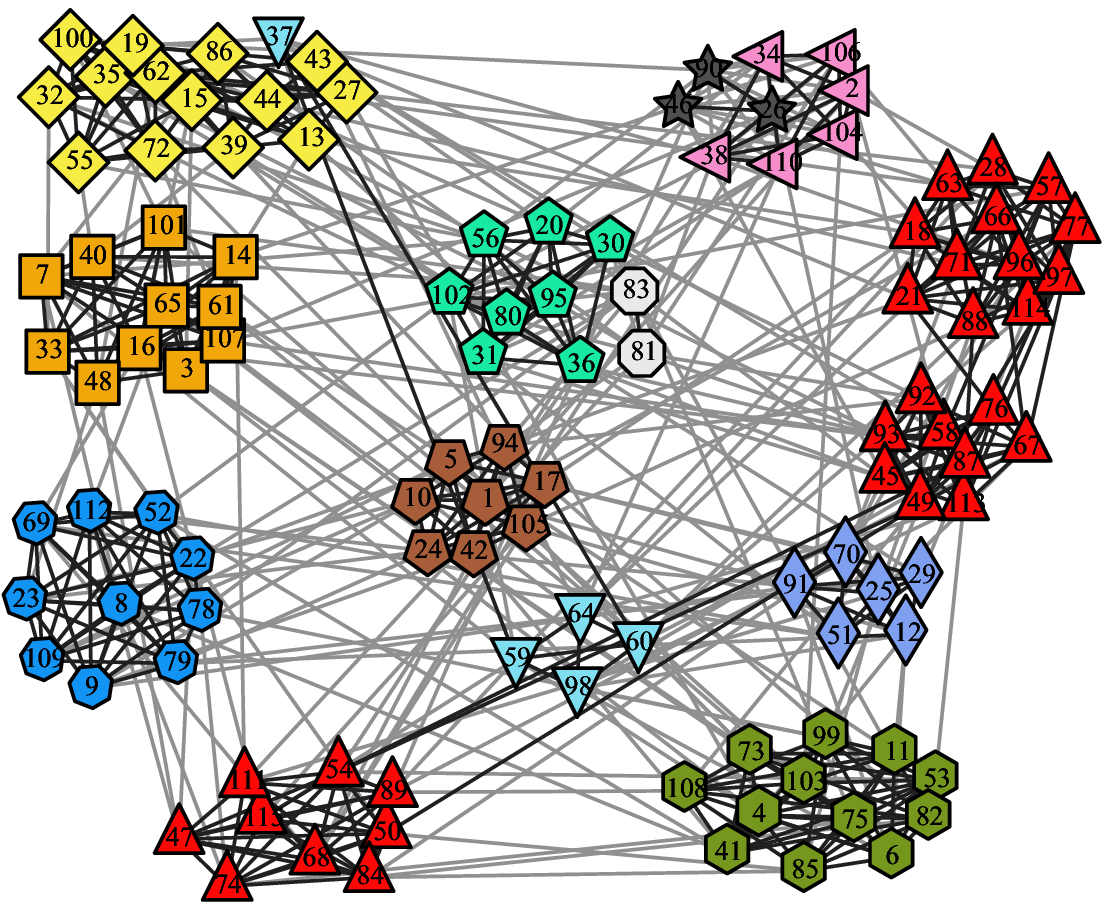}
}
\subfloat[\label{fig:football:newman2006}]{
	\includegraphics[scale=\figscale]{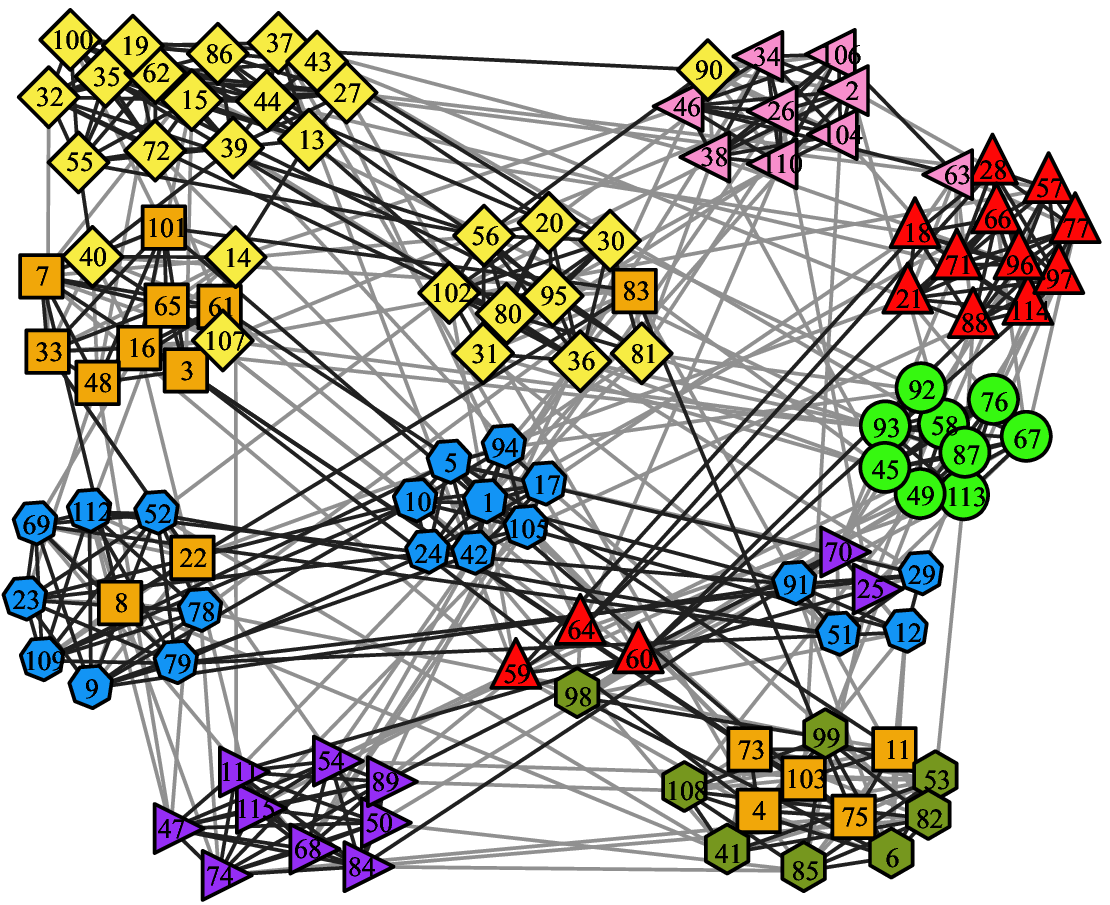}
}\\
\subfloat[\label{fig:football:infohiermap}]{
	\includegraphics[scale=\figscale]{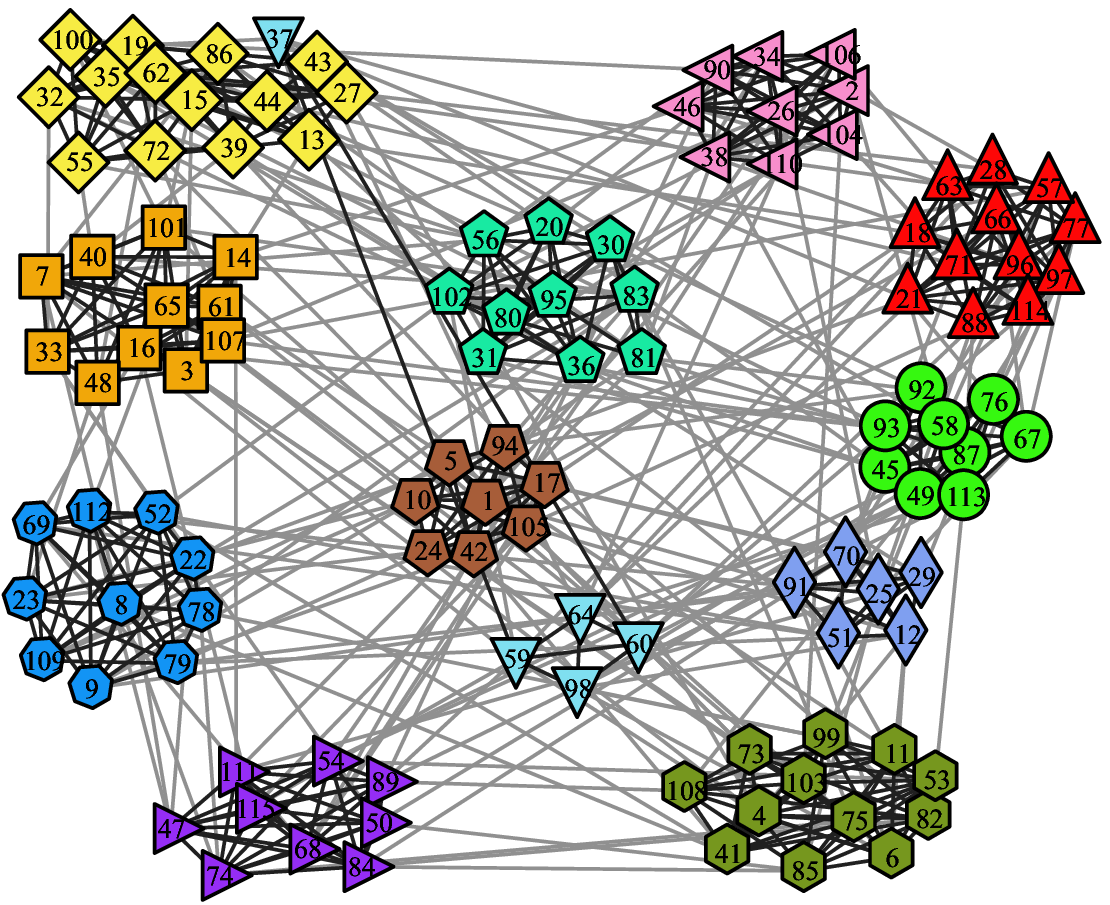}
}
\subfloat[\label{fig:football:without_sparsification}]{
  \includegraphics[scale=\figscale]{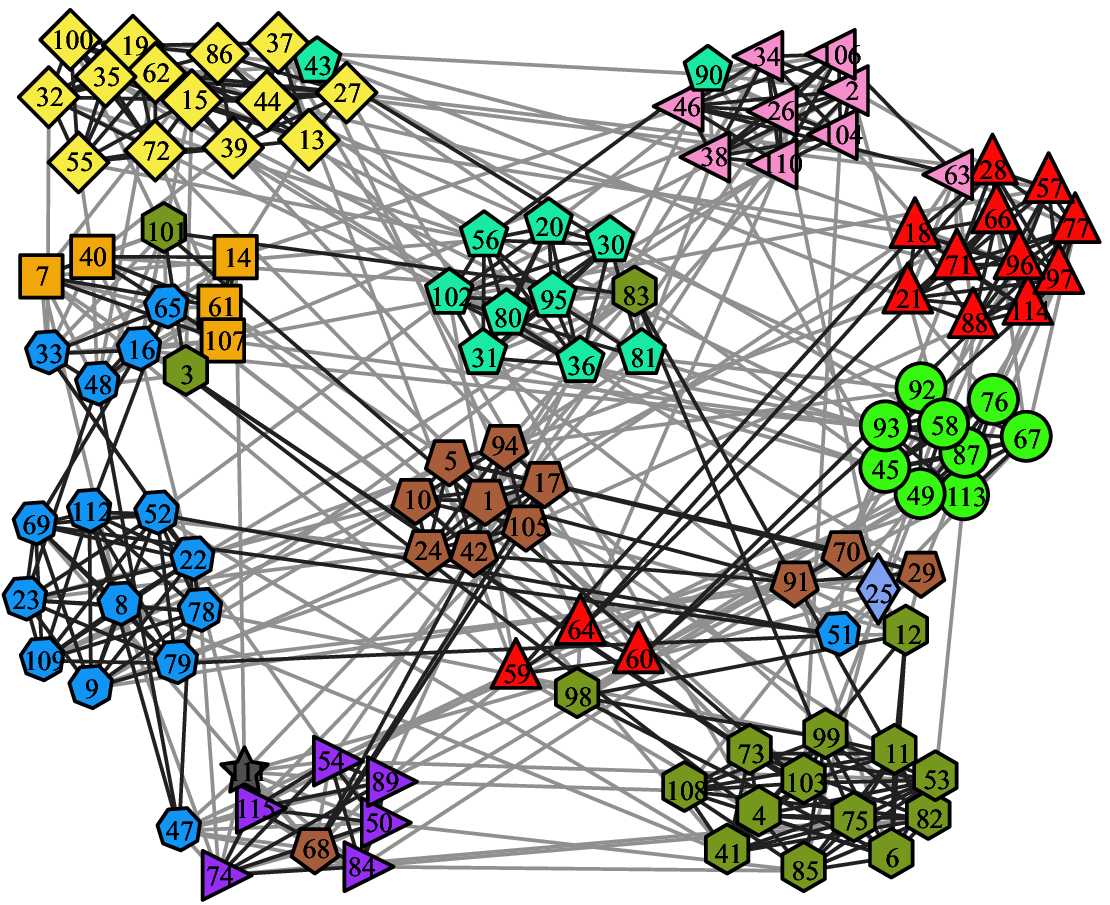}
}
\subfloat[\label{fig:football:proposal}]{
	\includegraphics[scale=\figscale]{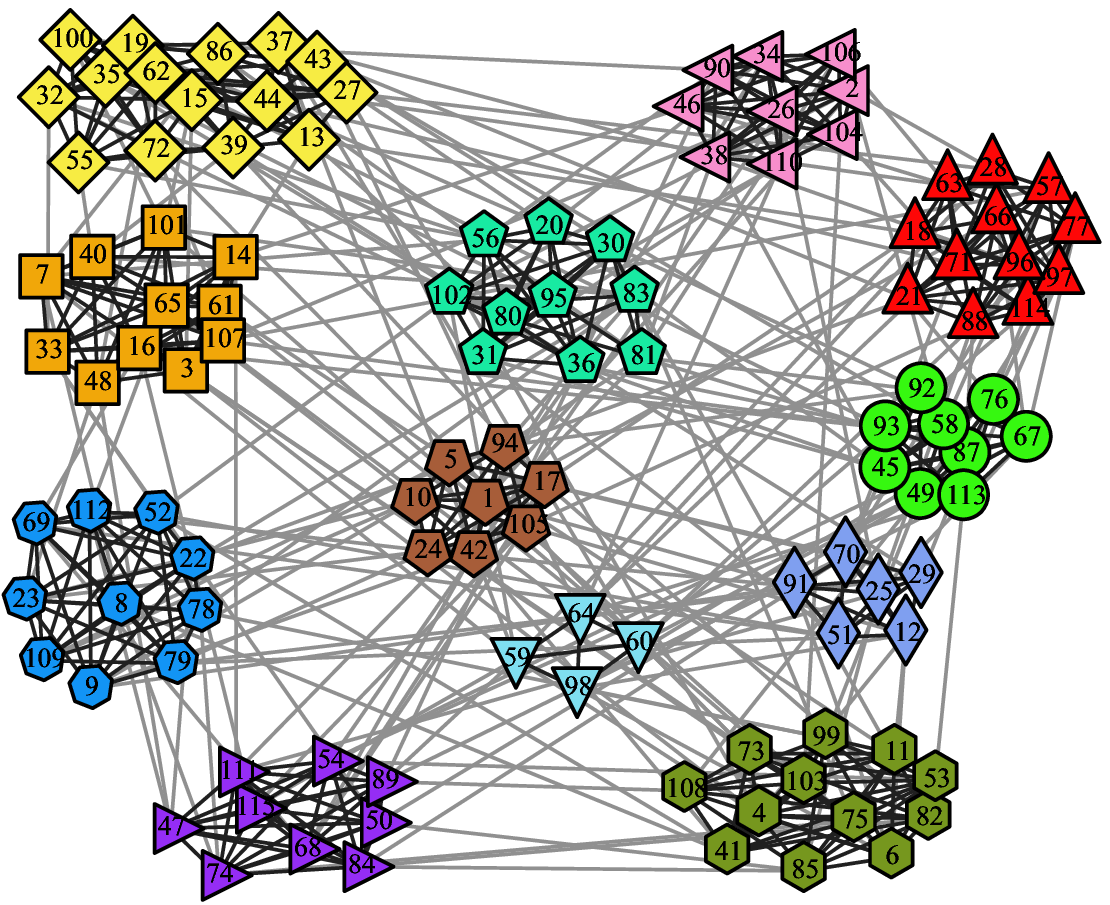}
}
\caption{\label{fig:football}\textbf{College football game schedule network.} \protect\subref{fig:football:ground_truth} The ground-truth community structure; \protect\subref{fig:football:spectral_clustering} The community structure extracted by the spectral clustering algorithm; \protect\subref{fig:football:newman2006} The community structure identified by Newman2006; \protect\subref{fig:football:infohiermap} The community structure uncovered by Infohiermap; \protect\subref{fig:football:without_sparsification} The community structure revealed by the lite version of the proposal; \protect\subref{fig:football:proposal} The community structure detected by the complete version of the proposed method.}
\end{figure*}

\textbf{College football game schedule network.} This network is the schedule of regular season Division I American college football games for year 2000 season. It is made up of 115 vertices and 613 edges, in which vertices represent teams and edges represent regular season games between the two teams they connect. The teams are divided into 12 ``conferences'', and games are more frequent between teams of the same conference than between teams of different conferences. Therefore, each conference is a natural community, and the ground-truth community structure is accordingly as shown in Figure \ref{fig:football:ground_truth}. Applying the comparison algorithms and the proposed method to this network, we achieved the resulting community structures presented in Figures \ref{fig:football:spectral_clustering}--\ref{fig:football:proposal}, correspondingly, and the values of the three metrics obtained on this network are filled in Table \ref{tab:comparisons} as well.

On this network, the spectral clustering algorithm tended to merge two or more communities into one, but to separate a small portion of vertices from some communities to form another communities (not only in the result presented here, but also in other results of the 20 runs of the algorithm on this network). For Newman2006, the quality of the result is quite poor as many vertices were classified into the incorrect communities wrongly. The similar result occurred for the lite version of the proposal, there exist too much misclassification of vertices. After sparsification, all mistakes were eliminated, the result of the complete version of our proposed method is identical to the ground truth. For Infohiermap, the extracted structure is almost identical with the ground-truth community structure, except for one vertex being misclassified. These results demonstrate that our proposed method performs the best again on this network.

\begin{table*}[htbp]
\centering
\caption{\label{tab:comparisons}\textbf{The comparisons of the 3 metrics.} We report the rank of each algorithm (in parentheses) on each metric per network, each score value in the last but one column is the average of the three metrics of each algorithm. The highest rank and the corresponding algorithm or method is shown in bold.}
\begin{tabular}{llllllc}\hline
network & algorithm & $Q$ & $A$ & $NMI$ & score & rank\\ \hline
karate & ground truth & 0.371 & 1.00 & 1.00 & &\\
       & spectral clustering & 0.313(6) & 0.912(4) & 0.646(6) & 4.667 & 6\\
       & Newman2006 & 0.393(2) & 0.618(6) & 0.677(5) & 4.333 & 5\\
       & Infohiermap & 0.402(1) & 0.824(5) & 0.699(4) & 3.333 & 4 \\
       & Newman2013 & 0.360(4) & 0.971(2) & 0.836(2)& 2.667 & 2 \\
       & lite  & 0.360(4) & 0.971(2) & 0.836(2) & 2.667 & 2 \\
       & \textbf{proposal} & 0.371(3) & 1.00(1) & 1.00(1) & \textbf{1.667} & \textbf{1}\\ \hline
dolphin & ground truth & 0.373 & 1.00 & 1.00 & & \\
        & spectral clustering & 0.379(6) & 0.984(1) & 0.889(1) & 2.667 & 5\\
        & Newman2006 & 0.491(2) & 0.484(6) & 0.449(6) & 4.667 & 6\\
        & Infohiermap & 0.525(1) & 0.581(5) & 0.566(5) & 3.667 & 4\\
        & \textbf{Newman2013} & 0.385(3) & 0.968(2) & 0.814(2) & \textbf{1.667} & \textbf{1}\\
        & \textbf{lite}  & 0.385(3) & 0.968(2) & 0.814(2) & \textbf{1.667} & \textbf{1}\\
        & \textbf{proposal} & 0.385(3) & 0.968(2) & 0.814(2) & \textbf{1.667} & \textbf{1} \\ \hline
Risk map & ground truth & 0.621 & 1.00 & 1.00 & & \\
        & spectral clustering & 0.589(3) & 0.833(3) & 0.818(3) & 3.000 & 3\\
        & Newman2006 & 0.547(5) & 0.762(4) & 0.723(4) & 4.333 & 4 \\
        & Infohiermap & 0.634(1) & 0.857(2) & 0.945(2) & 1.667 & 2\\
        & lite & 0.554(4) & 0.643(5) & 0.705(5) & 4.667 & 5\\
        & \textbf{proposal} & 0.631(2) & 0.976(1) & 0.956(1) & \textbf{1.333} & \textbf{1}\\ \hline
collaboration & ground truth & 0.739 & 1.00 & 1.00 & & \\
        & spectral clustering & 0.695(5) & 0.703(4) & 0.772(5) & 4.667 & 4\\
        & Newman2006 & 0.708(3) & 0.831(3) & 0.834(3) & 3.000 & 3 \\
        & Infohiermap$^{1st}$ & 0.651(6) & 0.636(5) & 0.764(6) & 5.667 & 6\\
        & Infohiermap$^{2nd}$ & 0.704(4) & 0.602(6) & 0.805(4) & 4.667 & 4\\
        & lite  & 0.734(2) & 0.924(2) & 0.895(2) & 2.000 & 2\\
        & \textbf{proposal} & 0.740(1) & 0.949(1) & 0.936(1) & \textbf{1.000} & \textbf{1}\\ \hline
football & ground truth & 0.601 & 1.00 & 1.00 & &\\
        & spectral clustering & 0.538(3) & 0.791(4) & 0.908(3) & 3.333 & 3\\
        & Newman2006 & 0.493(5) & 0.652(5) & 0.758(5) & 5.000 & 5 \\
        & Infohiermap & 0.600(2) & 0.991(2) & 0.989(2) & 2.000 & 2\\
        & lite  & 0.503(4) & 0.809(3) & 0.811(4) & 3.667 & 4\\
        & \textbf{proposal} & 0.601(1) & 1.000(1) & 1.000(1) & \textbf{1.000} & \textbf{1}\\
\hline
\multicolumn{7}{l}{\tabincell{p{0.925\textwidth}}{\footnotesize lite: the proposed repeated bisection algorithm without network sparsification; proposal: the complete version of our proposed method. Infohiermap$^{1st}$, Infohiermap$^{2nd}$: the first-level and the second-level community structures extracted by Infohiermap, respec\-tive\-ly.}}\end{tabular}
\end{table*}

At last, Let's make an analysis on the values of the three evaluation metrics, which have been recorded in Table \ref{tab:comparisons} in the procedure of experiments. From the perspective of the modularity, Infohiermap achieved the largest value 3 times (on the karate club network, the dolphin social network and the Risk map network, respectively), the complete version of our proposal acquired twice (on the scientist's collaboration network and the football game schedule network, respectively). Other algorithms have no chance to get the largest value on any one of the 5 networks. Considering from the perspective of the accuracy and $NMI$, except for being the second once (on the dolphin social network) only by a very small offset to the spectral clustering algorithm, our proposed method obtained the largest value on all 4 other networks steadily. Considering the meaning of the accuracy and $NMI$, these results suggest that the community structure extracted by our proposed method approaches the ground-truth community structure most. Furthermore, we attached a rank (the number in the parentheses) to each value on each metric per network, and calculated a score to rank the algorithms or methods totally by averaging the rank numbers of every algorithm. The final rank of every algorithm is listed in the last column of Table \ref{tab:comparisons}, which confirm that our proposed method performs much better than the comparison algorithms.

\section{Conclusion\label{conclusion} and future work}
In this paper, we proposed a novel spectral method to identify community structures from networks, which is a combination of a network-sparsification algorithm and a repeated bisection spectral community detection algorithm. First, some inter-community edges are removed by the sparsification algorithm to make the community structure more prominent, then the repeated bisection spectral algorithm extract the community structure accurately from the sparsified network. We have conducted extensive experiments on 5 real-world networks, and the experimental results show that our proposed method is superior to the comparison algorithm significantly.

The network sparsification algorithm is of great importance to our proposed method. To be frank, the strategy employed to remove some edges from the network in this paper is a bit too naive, the similarity threshold, $\theta$, is in fact a global parameter, so the network sparsification determine whether to remove an edge or not from the global perspective of the entire network, without considering any local property of any end vertex of the edge. Hence, some edges across communities but located in the region that the connection is relatively denser will not be removed, this might influence the quality of the result. And this might be the reason why there are still 6 vertices that are misclassified in the resulting community structure extracted by our proposed method from the scientist's collaboration network.

Therefore, although the network sparsification algorithm proposed in this paper does take its effect, we think that a sophisticated network sparsification strategy exploiting the local properties of edges, e.g., the densities of end vertices, will perform better. And network sparsification might be a research direction in the future not only for the need of community detection, but also for the demand of efficiency considering the larger and larger scales of networks.

\phantomsection
\addcontentsline{toc}{section}{\hspace{1em} Acknowledgments}
\ack{
  This work was partially supported by the Fundamental Research Funds for the Central Universities (grant Id: lzujbky-2014-54, lzujbky-2014-47).
}

\section*{References}
\phantomsection
\addcontentsline{toc}{section}{\hspace{1.4em}References}
\bibliographystyle{unsrt}
\bibliography{divisive_spectra}
 
\end{document}